\documentstyle[aps]{revtex}

\begin{document}

\title{The Singlet-Triplet Pseudo-Jahn-Teller Centers in Copper Oxides.}
\author{A.S. Moskvin, Yu.D. Panov.}
\address{Department of Physics, Ural State University, Ekaterinburg, 
Russia }

\maketitle
\begin{abstract}
One of the most exciting features of the hole centers $CuO_{4}^{5-}$ 
in doped cuprates is an unusually complicated ground state which is 
the result of the electronic quasi-degeneracy. An additional hole, 
doped to the basic $CuO_{4}^{6-}$ cluster with the $b_{1g}$ hole can 
occupy both the same hybrid $Cu3d-O2p$ orbital state resulting in a 
Zhang-Rice singlet $^1A_{1g}$ and the purely oxygen $e_u$ molecular 
orbital resulting in a singlet or triplet $^{1,3}E_u$ term with the 
close energies. We present detailed analysis of the 
(pseudo)-Jahn-Teller effect driven by the near-degeneracy within the 
$^1A_{1g},^{1,3}E_u$-manifold.  
\end{abstract}

\section{Introduction.}

Intuitive ideas concerning a specific role of Jahn-Teller ions
(centers, polarons) have been used as a starting point of the pioneer
investigations by K.A.Muller and J.G. Bednorz resulting in 1986 in the
outstanding discovery of the high-$T_c$ superconductivity 
\cite{Muller}.  Unfortunately, in the following years no breakthrough 
in understanding of this puzzling phenomenon occurred. In many 
respects such  situation is accounted for the underestimation of 
Jahn-Teller-effect and related phenomena that is typical for the 
conventional "metallic" approaches to the description of the 
electronic structure of the cuprates: namely these approaches underly  
the majority of the popular scenario's for the high-$T_c$ 
superconductivity. Additional argumentation of the opponents of the 
Jahn-Teller approach is based on the widespread opinion that the 
$CuO_4$ cluster with the doped hole forms a well-isolated spin and 
orbital singlet $^1A_{1g}$ (Zhang-Rice singlet).

Well-isolated Zhang-Rice singlet is a natural starting point for many  
model approaches including the well-known t-J-model. However, it does 
not  provide an explanation of wide set of unconventional physical 
properties of the cuprates, in particular, those associated with 
anomalous lattice and electron-lattice effects, and should be 
reconsidered and generalized.

A considerable number of  experimental data were accumulated which 
more or less directly give evidence in favor of existence of the 
copper-oxygen centers with near-degeneracy effects and anomalously 
strong electron-vibrational (or Jahn-Teller) correlations. More and 
more experimental data argue that a single band picture for the low 
energy excitations in the cuprates is inadequate.

An observation of the so-called mid-infrared (MIR) absorption bands in 
different $CuO_{4}$-cluster based oxides is one of the most impressive 
manifestations of near-degeneracy effects \cite{Moskvin1994} and, 
moreover, this phenomenon provides an important information about the 
energy spectrum and electronic structure within the ground state 
manifold. An appearance of the MIR bands with specific transformation 
of the absorption spectra in cuprates upon doping leads to a 
conjecture that the latter is accompanied by the sharp correlational 
decrease in the energy of the charge transfer transition 
$b_{1g}\rightarrow e_{u}$ that determines the fundamental absorption 
band for the parent oxides \cite{Rice1988}. Thus, ''parent'' 
absorption band shifts from rather usual position around 
$\sim 2\div3\,eV$ 
to the mid-infrared region forming the MIR band. An appearance 
of the MIR bands upon hole doping could be readily explained by 
assuming that additional hole doped to the basic $CuO_{4}$ cluster 
with the $b_{1g}$ hole can occupy both the same hybrid $Cu3d-O2p$ 
orbital state resulting in the Zhang-Rice singlet $^{1}A_{1g}$ and the 
purely oxygen $e_{u}$ molecular orbital resulting in the singlet or 
triplet $^{1,3}E_{u}$ term with the close energies. Then the 
MIR-absorption is determined by the allowed charge transfer transition 
$b_{1g}^{2}:{\,}^{1}A_{1g}\rightarrow b_{1g}e_{u}:{\,}^{1}E_{u}$ and 
represents the correlation analogue of the corresponding 
single-particle $b_{1g}\rightarrow e_{u}$ transition (see Fig.\ref{fig1}).

Perhaps, a detection of isolated PJT center would provide the direct 
manifestation of the validity of the Jahn-Teller conception. In this
connection the paper \cite{Yoshinari1996}  should be noted where the 
authors have performed NQR study of the isolated hole centers in
$La_{2}Cu_{0.5}Li_{0.5}O_4$. The results can be interpreted as 
convincing evidence of the quasi-degenerated singlet-triplet structure 
of the hole center. This conclusion is based on the following:

1. The authors have detected spin singlet ground state ($S=0$) and low
lying spin triplet state ($S=1$) with singlet-triplet separation
$\Delta_{ST} = 0.13\, eV$.

2. They observed anomalously weak temperature dependence of the 
relaxation rate at low temperatures that gives evidence of occurrence 
of the spinless multiplet structure in the $CuO_4$ cluster ground 
state.

3. They've found considerable spin contribution to the low temperature
relaxation indicating the simultaneous occurrence of the ground state
multiplet structure, sufficiently low singlet-triplet separation and
intrinsic singlet-triplet spin-orbital mixing.

4. They observed the relaxation inequivalence of the various $Cu$ 
sites, which is quite natural for the PJT centers in the conditions of 
the static PJT effect.

The Jahn-Teller hole centers like $CuO_4^{5-}$ with singlet-triplet
quasi-degeneracy within ground state have been observed by ESR 
spectroscopy in $LaSrAl_{1-x}Cu_{x}O_4$ which is isostructural to 
$La_{2-x}Sr_{x}CuO_4$ \cite{Ivanova}. Moreover, Yu. Yablokov et al. 
\cite{Yablokov1994} conjectured that doped hole in the copper-oxygen 
clusters occupies a purely oxygen $a_{2g}(\pi)$, or $b_{2u}(\pi)$ like 
orbitals. An important indication to the $O2p(\pi)$ nature of doped 
holes and, hence, to the occurrence of near-degeneracy for
configurations like $b_{1g}^{2}$ and $b_{1g}e_{u}(\pi)$ was obtained 
by Yoshinari \cite{Yo} and Martindale et al. \cite{Martindale} after 
the analysis of the $^{17}O$ Knight shift data and 
temperature-dependent anisotropy of the planar oxygen nuclear 
spin-lattice relaxation rate in $YBa_{2}Cu_{3}O_{6+x}$, respectively. 
Analogous conclusion could be drawn out of the comparative analysis of 
the temperature behavior for the nuclear spin-lattice relaxation rate 
in $La_{2-x}Sr_{x}CuO_4$ \cite{Walstedt}.  All this implies a 
complicated nature of the ground state manifold for the $CuO_4$ center 
with a significant mixing of the Zhang-Rice singlet and some other 
molecular term, which symmetry should be distinct from $^1A_{1g}$. 
This conclusion conflicts with the widespread opinion regarding the 
well isolation of the Zhang-Rice singlet.

An important argument in favor of vibronic nature for ground state of 
$CuO_4$ clusters with the participation of $e_u$-orbitals in the 123 
system was obtained after analysis of experimental data on EFG 
(electric field gradient) tensor for different nuclei in 123 system 
\cite{Master}: non-contradictory description of the data implies a 
considerable ($10\%$!) difference in electron density for $O(2)$ and 
$O(3)$ oxygens. It is unlikely that this result could be obtained 
without PJT effect.

These and many other results of resonance (ESR, NQR/NMR) experiments 
being precise local probes cast doubt on the validity of popular 
conceptions which are widely used as a starting point for the analysis 
of resonance and in broader sense for many other physical effects in 
cuprates.

A considerable number of the experimental indications of Jahn-Teller 
(vibronic) effects is associated with observation of lattice 
instabilities, ferroelectric \cite{ferro}, pyro- and piezoelectric 
\cite{pyro} anomalies, local static and dynamic distortions 
\cite{Egami}, various phonon anomalies and manifestation of 
substantial electron-phonon effects such as the line shift and Fano 
effect for the phonon modes associated with local PJT-active modes 
\cite{Pint,Fano,Plakida}. These observations and many other lattice 
effects are signatures of unconventional strong electron-lattice 
interaction at work in the cuprates \cite{Egami} with highly nonlinear 
and nonadiabatic intrinsic dynamics. Up to now these phenomena are 
often considered as convincing evidence in favour of the "structural" 
scenario for the high-$T_c$ superconductivity usually associated with 
polarons (pseudo-Jahn-Teller polarons) or bipolarons 
\cite{Egami,Bianconi}.  A number of the phonon anomalies could be 
associated with the vibronic pseudo-spin fluctuations or, in other 
words, with effects of the short-range cooperative Jahn-Teller 
ordering \cite{FTT}. These anomalies are linked with the specific 
points both inside the $BZ$ and on its boundary.

In our opinion the indirect evidence for Jahn-Teller nature of the 
$CuO_4$ centers with active role namely of the copper-oxygen hybrid 
$Q_{e_u}$ mode was displayed by the maximum entropy method (MEM) in 
$YBa_{2}Cu_{3}O_{6+x}$ at $x\sim 1$ \cite{MEM}. The authors observed 
characteristic squarish deformation of the nuclear density for $Cu$ 
atoms in the $CuO_2$ plane due to the anomalously strong anharmonic 
low temperature ($T=15 K$) motion presumably of vibronic nature.
Unconventional isotope effect and anomalous anisotropic pressure 
effect on $T_c$ in doped cuprates could also be associated with 
vibronic effects.

Thus, numerous experimental data show that Zhang-Rice model should be 
generalized with inclusion of near-degeneracy effects accompanied by 
PJT effect.

Many authors have treated JT (or PJT) effect in doped cuprates in a
rather general and various form both as a source of the local pairing 
\cite {Clou} and as a source of the unconventional physical properties
\cite{Bersuker1997,Kami,Mark,Mihailov}.  So, H. Kamimura \cite{Kami} 
proposed a mechanism of the HTSC due to the coherent bipolaron 
conduction induced by JT distortions. A dynamic Van Hove Jahn-Teller 
effect has been introduced by R.S. Markiewicz \cite{Mark}. The essence 
of the vibronic model by M. Georgiev et al., L. Mihailov et al. 
\cite{Mihailov} is existence of off-centered apex oxygens as local 
polarizable states due to PJT effect. Their model is one of the 
numerous so-called anharmonic models of the HTSC based on the account 
of anharmonic motion of apical oxygen \cite{Plakida}. However, it 
should be noted that recent studies by pulsed neutron scattering 
suggest that it is the in-plane $Cu$ site rather than the apical 
oxygen site that may be split into two positions \cite{Egami}. The 
absence of strong apical anomalies was stated earlier in 
Ref.\cite{Pint}.

In the most cases JT (PJT) effect has been considered within standard 
$E-e$-problem \cite{Bersuker1983} for $Cu^{2+}$ ion in octahedral 
environment (214 systems) or in square pyramidal environment
\cite{Bacci} (123 systems). Instead of standard $d_{x^{2}-y^{2}}$ , 
$d_{z^{2}}$ doublet some authors considered $d_{x^{2}-y^{2}}$ , 
$d_{xz}$, $d_{yz}$ manifold \cite{Ivanov}. However, all approaches 
originated from the assumption of near-degeneracy for some 
predominantly $Cu3d$-states do not agree with experimentally observed 
large gap ($\sim 1.5\, eV$) separating ground $d_{x^{2}-y^{2}}$ and 
any other $Cu3d$-states.  Besides, any JT model approach pretending to 
be universal should be originated from $CuO_4$ center as the only 
common element of the crystalline and electronic structure of all the 
cuprates.

A transformation of $CuO_4$ clusters into PJT centers upon hole or
electron doping to $CuO_2$ planes is a principal element of the so
called singlet-triplet PJT (ST-PJT) center model developed in \cite
{Moskvin1994,FTT,Moskvin1993,Moskvin1997,MOK,Moskvin1998,isotope}. In
addition to cuprates like $YBa_2Cu_3O_{6+x}$, $La_{2-x}Sr_xCuO_4$, 
$La_2CuO_{4+\delta}$ this model could be readily extended to a series 
of strongly correlated oxides like $(K,Ba)BiO_3$, $La_{1-x}Sr_xMnO_3$, 
$ La_2NiO_{4+\delta }$ including systems with the high-$T_c$ 
superconductivity and colossal magnetoresistance. Their unconventional 
properties reflect a result of  response of the system to nonisovalent 
substitution that stabilizes phases providing the most effective 
screening of charge inhomogeneity. These phases in oxides can involve 
novel unconventional molecular cluster configurations like Jahn-Teller 
$sp$-center \cite {Moskvin1993} with anomalous high local 
polarizability and multi-mode behavior.

The copper oxides based on $CuO_{4}$ clusters within this model are 
considered to be systems, which are unstable with regard to 
disproportionation reaction
\begin{equation}
2CuO_{4}^{6-} \rightarrow 
\left[ CuO_{4}^{5-}\right]_{JT}+\left[CuO_{4}^{7-}\right]_{JT}
\end{equation}
with formation of  system of polar (hole - $CuO_{4}^{5-}$ or electron 
- $CuO_{4}^{7-}$) pseudo-Jahn-Teller (PJT) centers. These centers are 
distinguished by the so-called local $S$-boson or two electrons paired
in the completely filled molecular orbital of the $CuO_{4}$-cluster. 
In other words, the novel phase can be considered to be  system of 
local spinless bosons moving in lattice of the hole PJT-centers 
$\left[ CuO_{4}^{5-}\right] _{JT}$ or the generalized quantum lattice 
bose-gas (or liquid) with boson concentration near $N_{B}=1/2$.

In a sense, this microscopic approach represents a particular 
generalization of Zhang-Rice model.

Though without detailed analysis of the 
$^{1}A_{1g},^{1,3}E_{u}-a_{1g}-b_{1g}-b_{2g}-e_u$ vibronic problem, 
this model approach has been successfully applied for qualitative and 
semi-quantitative description of many physical properties of cuprates: 
MIR absorption bands \cite{Moskvin1994}, isotope effect 
\cite{isotope}, static and dynamic magnetic properties 
\cite{Moskvin1998a}, local structure distortions \cite{local}, 
hyperfine coupling \cite{HFI}, phonon anomalies \cite{MOK}, neutron 
scattering \cite{Moskvin1998c}.  Further development of the model and, 
first of all, possibilities of quantitative predictions implies a 
detailed analysis of the 
$^{1}A_{1g},^{1,3}E_{u}-a_{1g}-b_{1g}-b_{2g}-e_u$ vibronic problem. 
Moreover, this problem is of great independent importance as a 
non-trivial example of multi-mode PJT effect.

Below, in our paper we consider in detail a vibronic structure of the 
isolated PJT center taking into account some effects associated with 
its singlet-triplet structure and spin-orbital coupling. First, in 
Section 2 a short consideration of the correlation driven near 
degeneracy effects for the $CuO_4$-clusters will be done.  Section 3 
contains a detailed analysis of the 
$^{1}A_{1g},^{1,3}E_{u}-a_{1g}-b_{1g}-b_{2g}-e_u$ vibronic problem 
within $CuO_4$ cluster including an adiabatic potential, vibronic 
states and tunnelling effects for three different regimes. Some 
effects of spin-orbital coupling within ($^1A_{1g},^{1,3}E_u$) 
manifold are considered in Section 4.

\section{Correlations and the near degeneracy effects for the $CuO_4$ 
clusters.}

At a glance the analysis of electronic structure and energy spectrum 
of the parent compounds such as $La_{2-x}M_xCuO_4$, $YBa_2Cu_3O_{6+x}$ 
at $x=0$ does not display any exotic peculiarities except 
quasi-two-dimensional antiferromagnetism determined by the strong 
 exchange interaction for the $b_{1g}(d_{x^2-y^2})$ holes in the 
"basic" $CuO_4^{6-}$ clusters. At the same time it is worth to pay 
attention to one important feature, namely to the exciton-band form of 
the fundamental absorption in the $1.5\div 3.0\ eV$ region strictly 
pronounced in the systems like $R_2CuO_4$, $YBa_2Cu_3O_{6+x}$, $CuO$ 
\cite{Moskvin1997}.

A peculiar character of this absorption connected with the allowed
charge-transfer transition $b_{1g}\rightarrow e_{u}$ between the
copper-oxygen $b_{1g}$-hybrid and the purely oxygen $e_{u}$-orbital 
provides evidence of the strongly correlated nature of the 
$e_{u}$-electrons with formal occurrence  of two types of the 
$e_{u}$-states with and without strong correlation. This peculiarity 
is associated with the maximal hole density occurred for oxygen ions 
just in the $e_{u}$-states of the $CuO_{4}$-cluster and can be easily 
 explained in the framework of the {\it non-rigid anionic background 
model} \cite{Hirsch1989}. This model introduces new correlation degree 
of freedom with two possible states of anionic background for the 
valent $O2p$-holes corresponding to two possible projections of the 
correlation pseudospin $s=1/2$ and is described by simplified 
Hamiltonian
\begin{equation}
H_{corr}=V_{1}\hat{\sigma}_{x}+V_{3}\hat{\sigma}_{z} ,
\end{equation}
where $V_{1,3}$ are two electronic operators for the valent states. 
Simple approximation used in \cite{Hirsch1989} conjectures the linear 
$n_{2p}$ -dependence for $V_{1,3}$, where $n_{2p}$ is the $O2p$-hole 
number. According to optical data \cite{Moskvin1997}, the correlation 
pseudospin splitting can achieve the value $~\sim 0.5\ eV$.

An increase in the $O2p$-hole concentration with hole doping 
 $CuO_4^{6-}\rightarrow CuO_4^{5-}$ results in a sharp increase of the 
$e_u$ correlation splitting in the hole $CuO_4^{5-}$-centers with 
transformation of the $b_{1g}\rightarrow e_u$ fundamental band to the 
high-energy $(b_{1g}^{2})\rightarrow (b_{1g} e_u)$ subband and the 
low-energy $b_{1g}^{2}\rightarrow b_{1g}^{*} e_u^{*}$ subband to be 
well known as MIR (mid-infrared) absorption band \cite{Moskvin1994}.

Fig.\ref{fig1} shows qualitatively the main results of the taking account 
of the considered ''$e_u$ correlations'' for energy spectrum of basic
$CuO_4^{6-}$ cluster and hole $CuO_4^{5-}$ cluster. It is important to
mention an appearance in our model of two types of the orthogonal (!) 
molecular orbitals; for instance $b_{1g}$ (upper correlation sublevel) 
and $b_{1g}^{*}$ (lower correlation sublevel) states, $e_u$ and 
$e_u^{*}$.

Thus, we come to a conclusion about a near-degeneracy for two 
configurations $ b_{1g}^{*\ 2}$ and $b_{1g}^{*}e_u^{*}$ with 
$b_{1g}^{*}$ and $e_u^{*}$ being the lowest correlation sublevels. 
This result does not drastically changes with taking account of an 
electrostatic interaction $V_{ee}$ between two holes. Moreover, just 
the $V_{ee}$ contribution was considered earlier \cite {Moskvin1994} 
as a main reason for a near-degeneracy for the $^1A_{1g}$ and 
$^{1,3}E_u$ terms formed by $b_{1g}^{*\ 2}$ and $b_{1g}^{*}e_u^{*}$ 
configurations. So, the both correlation effects lead to a 
near-degeneracy in  ground state of the hole center $CuO_4^{5-}$.

Unusual properties of the $(^1A_{1g},^{1,3}E_u)$ manifold involving 
terms distinguished by the spin multiplicity, parity and orbital 
degeneracy provides unconventional behavior for the hole center 
$CuO_4^{5-}$ with active interplay of various modes. As an extremely 
important result one should note that $E_u$-doublet has a nonquenched 
Izing like orbital moment that can be directed only along the 
$C_4$-axis.

A near degeneracy within $(^1A_{1g},^{1,3}E_u)$ manifold can lead to
conditions for the pseudo-Jahn-Teller effect \cite{Bersuker1983} with
anomalously strong electron-lattice correlations with active local
displacements modes of the $Q_{e_u}$, $Q_{b_{1g}}$ and $Q_{b_{2g}}$
types. It should be noted here that the $Q_{e_u}$ modes are the only 
hybrid copper-oxygen modes, while the $Q_{b_{1g}}$ and $Q_{b_{2g}}$ 
modes are the pure oxygen ones.

\section{ Vibronic coupling for the $CuO_{4}^{5-}$ centers.}

\subsection{ Adiabatic potential.}

Below we'll make use the notation $|SM_{S}\Gamma \gamma \rangle$ for
basis wave functions from the $(^{1}A_{1g},^{1,3}E_{u})$ manifold (see
Fig.\ref{fig1}). Here, $S$ $(=0,1)$, $M_{S}$ are the total spin and its 
projection, $\Gamma \gamma $ ($=A_{1g}$, $E_{u}^{x}$, $E_{u}^{y}$) 
labels the irreducible representation of symmetry group $D_{4h}$ for 
the $CuO_{4}$ center and its row, respectively, indicating the 
transformation properties of the orbital functions. Subsequently, we 
restrict ourselves with the linear vibronic coupling within the 
$(^{1}A_{1g},^{1,3}E_{u})$ manifold with the JT-active vibrational 
coordinates of the $a_{1g}$, $b_{1g} $, $b_{2g}$, $e_{u}$ symmetry.

Vibronic coupling for the isolated singlet or triplet $^{1,3}E_{u}$ 
term has a well known for the $E-b_{1}-b_{2}$-problem 
\cite{Bersuker1983} form diagonal in $S$ and $M_{S}$
\begin{equation}
\left(
\begin{array}{cc}
V_{b_{1g}}^{\tau }Q_{b_{1g}} & V_{b_{2g}}^{\tau }Q_{b_{2g}} \\
V_{b_{2g}}^{\tau }Q_{b_{2g}} & -V_{b_{1g}}^{\tau }Q_{b_{1g}}
\end{array}
\right) ,\quad \tau =\ ^{1}E_{u},\ ^{3}E_{u}.
\end{equation}

Singlet terms $^{1}E_{u}$ and $^{1}A_{1g}$ interact due to linear
vibronic coupling
\begin{equation}
\left\langle 00A_{1g}|\hat{V}_{vib}|00E_{u}^{i}\right\rangle
=\sum_{e_{u}}V_{e}Q_{e_{u}^{i}}
\end{equation}
determined by active vibrational coordinates 
$Q_{e_{u}^{x}},Q_{e_{u}^{y}}$.  For the $CuO_{4}$ cluster there are 
 three normal coordinates with $e_{u}$ symmetry, however, below we 
restrict ourselves with the choice of one active $e_{u}$ vibration 
with an appropriate linear vibronic coupling constant $V_{e}$.

The sum of elastic energy $\hat{V}_{Q}$, electronic Hamiltonian 
$\hat{V}_{el} $, and vibronic Hamiltonian $\hat{V}_{vib}$ for the 
singlet $S=0$ spin manifold ($^{1}A_{1g}$,$^{1}E_{u}$) with bare 
separation $\Delta_{AE}$ (see Fig.\ref{fig1}) could be written as:
\begin{equation}
\hat{U}(Q)=\sum_{i}\frac{\omega_i^2 Q_i^2}{2}\cdot \hat{I}+\left(
\begin{array}{ccc}
-\Delta +V_{z}Q_{z} & V_{e}Q_{x} & V_{e}Q_{y} \\
V_{e}Q_{x} & V_{\alpha }Q_{\alpha } & V_{\beta }Q_{\beta } \\
V_{e}Q_{y} & V_{\beta }Q_{\beta } & -V_{\alpha }Q_{\alpha }
\end{array}
\right) ,
\end{equation}
where the indices both for the coupling coefficient and normal 
coodinates are defined as follows: $a_{1g}\rightarrow z$, 
$e_{u}^{x}\rightarrow x$, $e_{u}^{y}\rightarrow y$, 
$b_{1g}\rightarrow \alpha $, $b_{2g}\rightarrow \beta $, and 
$\Delta =\Delta_{AE}-V_{z}q_{z}^{(0)}$, 
$V_{z}=V_{z}^{A_{1g}}-V_{z}^{E_{u}}$, 
$q_{i}^{\left( 0\right) }=V_{i}/\omega _{i}^{2}$.

An important information for the PJT center could be obtained with
examination of the adiabatic potential (AP) surfaces $\varepsilon 
(Q)$, which are the roots of characteristic equation for $\hat{U}(Q)$. 
In our case this is reduced to a cubic equation with an extremely 
complicated expression for the roots. The coordinates of the minima 
$Q^{0}$, energy and structure of electronic wave function at $Q^{0}$ 
and curvature of energy surface near $Q^{0} $ can be obtained by the 
Opik-Pryce method \cite{OpikPryce1957}, where an eigenvalue problem is 
treated only for extremal points of the AP. It should be noted that 
the Opik-Pryce method does not permit to find specific points of the 
AP without definite values of the derivative, however namely such a 
situation occurs for the upper sheets of the AP. The type of minima 
can be derived from curvature analysis for $\varepsilon (Q)$ near 
$Q^{0}$.

The relations between the extremum coordinates and the wave
function coefficients are
\begin{eqnarray}
&& Q_{z}^{0} = -q_{z}^{\left( 0\right) }z^{2},\quad
Q_{x}^{0}=-2q_{e}^{\left( 0\right) }zx,\quad Q_{y}^{0}=-2q_{e}^{\left(
0\right) }zy, \\
&& Q_{\alpha }^{0} = -q_{\alpha }^{(0) }\left(
x^{2}-y^{2}\right) ,\quad Q_{\beta}^0 =-2 q_{\beta}^{(0)} x y,
\nonumber
\end{eqnarray}
where we have denoted the coefficients of decomposition of the 
electronic wave function as $z$, $x$ and $y$ for 
$\left| 00A_{1g}\right\rangle ,
\left|00E_{u}^{x}\right\rangle $ and 
$\left| 00E_{u}^{y}\right\rangle $,
respectively. The eigenvalue problem for $V_{vib}$ in the 
AP extremum points  is given by:
\begin{equation}
\left(
\begin{array}{ccc}
-\Delta -2E_{JT}^{z}\ z^{2} & -4E_{JT}^{e}\ zx & -4E_{JT}^{e}\ zy \\
-4E_{JT}^{e}\ zx & -2E_{JT}^{\alpha }\ \left( x^{2}-y^{2}\right) &
-4E_{JT}^{\beta }\ xy \\
-4E_{JT}^{e}\ zy & -4E_{JT}^{\beta }\ xy & 2E_{JT}^{\alpha }\ \left(
x^{2}-y^{2}\right)
\end{array}
\right) \left(
\begin{array}{c}
z \\
x \\
y
\end{array}
\right) =\lambda \left(
\begin{array}{c}
z \\
x \\
y
\end{array}
\right) ,  \label{system}
\end{equation}
where $E_{JT}^{i} = \frac{1}{2} V_{i} q_{i}^{(0)}=
V_{i}^{2}/2 \omega_{i}^{2}$ is the specific JT energy.

The system (\ref{system}) complemented with a normalization condition 
$x^{2}+y^{2}+z^{2}=1$ has 13 solutions, which are listed in the Table 
\ref{table1}. Also the expressions for the quadratic form of the AP surface near the 
extremal points are given. These solutions could be 
divided into three groups.

1. The first group NJT (non-JT) contains the only solution. Electronic 
part of wave function in the extremal point is pure 
$\left|00A_{1g}\right\rangle$.  The NJT-extremum is the minimum on 
lower sheet of AP, if $b<0$ $(b=-\Delta +4E_{JT}^{e}-2E_{JT}^{z})$ or
 on the upper sheet, if $(\Delta +2E_{JT}^{z})<0$. In both cases the 
weak pseudo-Jahn-Teller effect takes place, when due to weakness of 
the vibronic coupling in comparison with the bare separation of 
electronic levels $\Delta $, there are no low-symmetry distortions of 
the $CuO_{4}$ cluster, and only the renormalization of the local 
$e_{u}$ vibration frequency occurs:
\begin{equation}
\left( \tilde{\omega}_{e}^{\left( NJT\right) }\right) ^{2}=\omega
_{e}^{2}\left( 1-\eta _{z}\right) ,\quad \eta _{z}=
\frac{4E_{JT}^{e}}{\Delta+2E_{JT}^{z}}\ .
\end{equation}

2. The second group JT$_{i}$ ($i=\alpha ,\beta $) contains four 
solutions, which are similar to results of the well-known 
$E-b_{1}-b_{2}$-problem.  The wave function at the extremum points is 
a pure $E_{u}$ superposition. In further analysis the rhombic mode 
with larger JT energy will be called the ''strong'' ($\sigma $) mode, 
and that with smaller JT energy will be called the ''weak'' 
($\sigma^{\prime }$) one: 
$E_{JT}^{\sigma } > E_{JT}^{\sigma^{\prime }}$,  
$(\sigma ,\sigma^{\prime } = \alpha ,\beta )$; for the 
$E-b_{1}-b_{2}$-problem the $E_{JT}^{\sigma }$ is the JT stabilization 
energy. Among four extremum points only the JT$_{\sigma }$ pair will 
correspond to minima. The minima are located on the lower sheet of the 
AP, if $a_{\sigma }<0$ $(a_{\sigma }=\Delta 
+4E_{JT}^{e}-2E_{JT}^{\sigma })$, and on the middle one, if 
$(\Delta-2E_{JT}^{\sigma })>0$. The JT$_{\sigma ^{\prime }}$ pair 
represents saddle points. The wave functions at the minima are 
orthogonal each other. If 
$E_{JT}^{\sigma }=E_{JT}^{\sigma ^{\prime }}$, the 
equipotential continuum of minima, or trough, exists. The $\sigma$ 
mode frequency does not vary, and that of $\sigma ^{\prime }$ mode is 
renormalized due to vibronic coupling:

\begin{equation}
\tilde{\omega}_{\sigma }^{2} = \omega _{\sigma }^{2}\ ,\quad 
\tilde{\omega}_{\sigma ^{\prime }}^{2}=
\omega _{\sigma ^{\prime }}^{2}\left( 1-\lambda
_{\sigma }\right) ,\quad 
\lambda _{\sigma } = 
\frac{E_{JT}^{\sigma^{\prime}}}{E_{JT}^{\sigma }}\ .
\end{equation}

It should be noted here that a type ($B_{1g}$ or $B_{2g}$) of the 
ground JT mode is of principal importance for the physics of the 
copper oxides.  This is determined by the competition of vibronic 
parameters for the $Cu3d-O2p$ and $O2p-O2p$ bonds minimizing the 
$B_{1g}$ and $B_{2g}$ modes, respectively.

For the $e_{u}$ vibrations the JT$_{i}$ solutions correspond to the 
weak PJT effect: only renormalization of local $e_{u}$ vibration 
occurs. If $\alpha $ mode is strong, the coordinates of minimum 
determine the rhombic distortion of the $CuO_{4}$ cluster along $x$- 
or $y$ -direction (Fig.\ref{fig2}). Accordingly, the softening of the 
$e_{u}^{x}$ or $ e_{u}^{y}$ vibration occurs; but as the both minima 
are equivalent, the frequencies of local $e_{u}$ modes remain twice 
degenerate. In a case of ''strong'' $\beta $ mode with rectangular 
distortion of the $CuO_{4}$ cluster, the softening of that of $e_u$ 
modes occurs, which co-directs to the cluster distortion. In all cases 
the expression for the renormalized local $ e_{u}$ mode frequency is 
written as:  
\begin{equation} 
\left( \tilde{\omega}_{e}^{\left(JT\right) }\right) ^{2} = 
\omega_e^2 \left( 1-\kappa_{\sigma}\right) \ ,\quad 
\kappa_{\sigma } = \frac{ 4E_{JT}^{e}}{-\Delta+2E_{JT}^{\sigma }}\ .  
\end{equation}

3. The third group PJT$_{i}$ ($i=\alpha ,\beta $) includes eight 
solutions and corresponds to the most complicated case of the strong
pseudo-Jahn-Teller effect. In this case the wave functions at the 
extremum points are the $A_{1g}-E_{u}$ hybrid states, and coefficients 
of superposition depend on the bare splitting $\Delta $ and the JT 
energies.  The four from the eight PJT$_{\sigma }$ extrema ($\sigma$ 
is the ''strong'' rhombic mode) are minima, if $a_{\sigma }>0$ and 
$\dot{b}>0$. All the PJT$_{\sigma }$ minima are equivalent and 
allocated on the lower sheet of the AP. An arrangement of minima in  
space of the normal coordinates of the $CuO_{4}$ cluster is 
schematically shown in Fig.\ref{fig3}. The wave functions at the minima 
are not orthogonal each other, that is a characteristic feature of strong 
pseudo-Jahn-Teller effect. The four PJT$_{\sigma^{\prime }}$ extrema 
($\sigma ^{\prime }$ is the ''weak'' rhombic mode) are saddle points. 
A specific case of degeneracy for the JT energies of rhombic modes 
should be examined separately.

The rhombic distortion of the $CuO_{4}$ cluster at the minimum points 
with the nonzero plane quadrupole moment is accompanied by the 
 co-directed $e_{u}$ distortion with the electric dipole moment. The 
possible PJT$_{\sigma}$ distortions are shown in Fig.\ref{fig2}.

Close to minimum $M$ the equipotential surface of the quadratic form 
$\varepsilon (Q-Q^{(0,M)})$ is a five-dimensional ellipsoid with center
located at $Q^{(0,M)}$ and principal axes to be turned with regard to 
basic ones. The $\sigma$ mode at minimum is mixed with co-directed 
$e_u$ mode and $a_{1g}$ mode giving rise to three local hybrid modes. 
The $\sigma^{\prime}$ mode is mixed with the second $e_u$ mode 
giving rise to two local hybrid modes. The mixing coefficients are 
proportional to the appropriate vibronic coupling constants.

In a case, when $V_{z}=0$, the frequencies of normal local hybrid modes
are written as:
\begin{equation}
\left( Q_{\sigma ^{\prime }},\tilde{Q}_{1}\right) 
:\qquad \omega _{\pm }^{2}=
\frac{A+B}{2}\pm \sqrt{\left( \frac{A-B}{2}\right) ^{2}+C^{2}}\ ,
\label{freq1}
\end{equation}
where
\begin{eqnarray}
&&A=\omega _{e}^{2}\left( 1-\upsilon _{\sigma }^{2}\right) ,\ 
B=\omega_{\sigma ^{\prime }}^{2}\left( 1-\rho _{\sigma }^{2}\right) ,\ 
C=\omega_{e}\omega_{\sigma^{\prime }}
\upsilon_{\sigma }\rho_{\sigma},  \nonumber
\\
&&\upsilon _{\sigma }=\sqrt{\frac{E_{JT}^{e}\ a_{\sigma }}{E_{JT}^{e}\
a_{\sigma }+E_{JT}^{\sigma }\ b}},\quad \rho _{\sigma }=
\sqrt{\frac{E_{JT}^{\sigma ^{\prime }}\ b}{E_{JT}^{e}\ 
a_{\sigma}+E_{JT}^{\sigma }\ b}},
\nonumber
\end{eqnarray}
and
\begin{equation}
\left( Q_{\sigma },\tilde{Q}_{2}\right) 
:\qquad \omega _{\pm }^{\prime \ 2}=
\frac{D+E}{2}\pm \sqrt{\left( \frac{D-E}{2}\right) ^{2}+F^{2}}\ ,
\label{freq2}
\end{equation}
where
\begin{eqnarray}
&&D=\omega _{e}^{2}\left( 1-\nu _{\sigma }^{2}\right) ,\ 
B=\omega _{\sigma}^{2}\left( 1-\mu _{\sigma }^{2}\right) ,\ 
F=\omega _{e}\omega_{\sigma }\nu_{\sigma }\mu_{\sigma },  \nonumber \\
&&\mu _{\sigma }=\sqrt{\frac{E_{JT}^{\sigma }\ 
a_{\sigma }b}{E_{JT}^{e}\
\left( a_{\sigma }+b\right) ^{2}}},\quad 
\nu _{\sigma }=\frac{a_{\sigma }-b}{a_{\sigma }+b}.  \nonumber
\end{eqnarray}

For the minima 1 and 3 $\tilde{Q}_{1}=Q_{x}$, if $\sigma =\alpha $ 
($Q_{1}$, if $\sigma =\beta $), $\tilde{Q}_{2}=Q_{y}$, 
if $\sigma = \alpha $ 
($Q_{2}$, if $\sigma =\beta $); for minima 2 and 4 
$\tilde{Q}_{1}=Q_{y}$, if $\sigma =\alpha $ ($Q_{2}$, 
if $\sigma =\beta $), 
$\tilde{Q}_{2}=Q_{x}$, if $\sigma =\alpha $ ($Q_{1}$, if 
$\sigma = \beta$). Due to an equivalence of minima all local 
frequencies coincide.

The type of minima on the lower AP sheet
will be determined mainly by following quantities:
\begin{equation}
a_{\sigma }=\Delta +4E_{JT}^{e}-2E_{JT}^{\sigma },\quad 
b=-\Delta
+4E_{JT}^{e}-2E_{JT}^{z}\
\end{equation}
in the following way:

a) NJT, if $a_{\sigma } > 0$ and $b < 0$;

b) JT$_{\sigma }$, if $a_{\sigma } < 0$ and $b > 0$;

c) PJT$_{\sigma }$, if $a_{\sigma } > 0$ and $b > 0$;

d) NJT and JT$_{\sigma }$, if $a_{\sigma } < 0$ and $b < 0$.

The diagram of states of the lower AP sheet in  space of parameters 
$\Delta $, $E_{JT}^{\sigma }$ and $E_{JT}^{e}$ at constant value of 
$E_{JT}^{z,0}$ is shown in Fig.\ref{fig4}. A cross-section of the parameter 
space for constant value of $E_{JT}^{e}$ is shown in Fig.\ref{fig5}a. If 
$\Delta \rightarrow +\infty $, the lowest $^{1}A_{1g}$ level is well 
isolated and the lower AP sheet  has a trivial NJT minimum. In 
contrast, at $\Delta \rightarrow -\infty $ the $^{1}E_{u}$ term 
becomes the lower one and usual $E-b_{1}-b_{2}$-problem with two 
JT$_{\sigma }$ minima on the lower AP sheet occurs.  When coming 
together, the $^{1}A_{1g}$ and $^{1}E_{u}$ terms are mixed by the 
$e_{u}$ mode, and the $e_{u}$ frequency is renormalized. This is 
accompanied by a formation of four PJT$_{\sigma }$ minima or three 
(NJT+JT$_{\sigma }$) minima depending on magnitude of 
$E_{JT}^{\sigma}$. These two possibilities correspond 
to lines $E_{JT}^{\sigma }=E_{JT}^{\sigma ,1}$ and 
$E_{JT}^{\sigma} = E_{JT}^{\sigma ,2}$ in Fig.\ref{fig5}a.

With a motion along the line $E_{JT}^\sigma =E_{JT}^{\sigma ,1}$ from 
$-\infty $ up to $+\infty $ the curvature of the JT$_\sigma $ minima 
 along $e_u$ directions decreases up to zero 
on the line $a_\sigma=0$. At this point 
the JT$_\sigma $ minima transform into saddle 
points with simultaneous appearence of the PJT$_\sigma $ minima (one 
of the JT$_\sigma $ minima splits along $Q_1$, another one splits 
along $Q_2$ direction). Moving from the line $a_\sigma =0$ up to $b=0$ 
the magnitude of rhombic distortion decreases up to zero on the line 
$b=0$. The magnitude of $e_u$ coordinates of minima firstly increases, 
then decreases up to zero, reaching a maximum on the line 
$\Delta =-E_{JT}^\sigma +E_{JT}^z$. The $Q_z$ 
coordinate increases linearly ($\sim a_\sigma $) from 
zero up to $-q_z^{\left( 0\right) }$ on the line 
$b=0$.  Thus, the four PJT$_\sigma $ minima transform into one NJT 
minimum. Further, with $\Delta \rightarrow +\infty $ the $A_{1g}$ and 
$E_u$ terms interact weaker and 
$\tilde{\omega }_e\rightarrow \omega_e$.

The PJT$_\sigma $ minima occur, if the $e_u$ mode driven 
interaction of the $A_{1g}$ and $E_u$ terms is rather effective. 
When $E_{JT}^\sigma>4E_{JT}^e-E_{JT}^z$ 
(for example $E_{JT}^\sigma =E_{JT}^{\sigma ,2}$), the 
$\sigma$ mode driven interaction becomes more effective. 
In this case at $b=0 $ the NJT minimum appears on the lower 
AP sheet  together with JT$_\sigma$ minima. All three minima 
have the same energy on the line $\Delta= -E_{JT}^\sigma+E_{JT}^z$. 
Further, at $\Delta \rightarrow +\infty $ the JT$_\sigma$ 
minima become more flat, without varying their coordinates, and at
$a_\sigma = 0$ only the NJT minimum remains.

In Fig.\ref{fig5}b the diagram of the upper AP sheets  is shown.

\subsection{Tunnel splitting.}

The most complicated case of strong pseudo-Jahn-Teller effect can be
treated in a framework of the tunnel Hamiltonian 
\cite{Bersuker1962},when the localized vibrations at the AP minima are 
considered with taking account of the inter-well tunneling. It is 
supposed, that the minimum depth is larger than the typical phonon 
energy, hence the tunnel frequency is rather small.

We consider the minimization problem for the total energy functional 
$E\left[ \Psi \right] =
\left\langle \Psi \left| \hat{H}\right| \Psi\right\rangle $ 
with Hamiltonian
\begin{equation}
\hat{H}=\hat{T}_{Q}+\hat{V}_{el}+\hat{V}_{Q}+\hat{V}_{vib}\ ,
\label{Tunnel_Hamiltonian}
\end{equation}
where $\hat{T}_{Q}$ is  kinetic energy of nuclei,
$\hat{V}_{el}$ is electronic energy operator,
$\hat{V}_{Q}$ is elastic energy, $\hat{V}_{vib}$ is vibronic 
Hamiltonian. The $\Psi$ is written as
\begin{equation}
\Psi =\sum_{M=1}^{4}c_{M}\ 
\varphi _{M}^{\left( \sigma \right) }\ \chi
_{M}^{\left( \sigma \right) }\ ,
\end{equation}
where $\varphi _{M}^{\left( \sigma \right) }$ 
and $\chi _{M}^{\left( \sigma\right)}$ 
are electronic (see Table \ref{table1}) and vibrational wave functions,
respectively, centered at the minimum 
PJT$_{\sigma }^{\left( M\right)}$.
The ground state vibrational wave function 
$\chi _{M}^{\left( \sigma \right)}$ has a form
\begin{equation}
\chi _{M}^{\left( \sigma \right) }=\prod_{k=1}^{5}\left( \frac{\omega
_{k}^{M}}{\pi }\right) ^{\frac{1}{4}}\exp \left\{ -\frac{1}{2}\omega
_{k}^{M}\left( \sum_{j=1}^{5}U_{jk}^{M}
\left( Q_{j}-Q_{j}^{\left( 0,M\right)
}\right) \right) ^{2}\right\} \ ,
\end{equation}
where $\left( \omega _{k}^{M}\right) ^{2}$ are the eigenvalues of 
a matrix of the quadratic form 
$\varepsilon \left( Q-Q^{\left( 0,M\right) }\right) $, $U^{M}$
is  matrix for the unitary transformation to principal axes for
$\varepsilon \left( Q-Q^{\left( 0,M\right) }\right) $. 
If $V_{z}=0$, the frequencies $\omega _{k}^{M}$ are equal to 
$\omega_{\pm }$ Eq. (\ref{freq1}),
$\omega_{\pm }^{\prime }$ Eq. (\ref{freq2}) and $\omega_{z}$,
respectively.

The variation of the energy functional $E\left[ \Psi \right] $ gives

\begin{equation}
H\ \vec{c}=E\ S\ \vec{c},  \label{tunnel system}
\end{equation}
where $H$ and $S$ are the Hamiltonian Eq. (\ref{Tunnel_Hamiltonian}) 
and overlap matrix, respectively. The solutions of the system Eq. 
(\ref{tunnel system}) give the tunnel states and the tunnel energy 
levels, respectively.

Due to equivalence of the PJT$_{\sigma }$ minima the $H$ and $S$ 
matrices include only three types of the non-zero matrix elements:

a) the diagonal matrix elements:

$\left\langle \varphi _{M}^{\left( \sigma \right) }\ 
\chi _{M}^{\left(
\sigma \right) }\left| \hat{H}\right| 
\varphi _{M}^{\left( \sigma
\right) }\ \chi _{M}^{\left( \sigma \right) }\right\rangle = H,
\quad
\left\langle \varphi _{M}^{\left( \sigma \right) }\ 
\chi _{M}^{\left( \sigma
\right) }|\varphi _{M}^{\left( \sigma \right) }\ 
\chi _{M}^{\left( \sigma
\right) }\right\rangle = 1$;

b) the non-diagonal matrix elements for the states with different both
dipole and quadrupole moments:

$\left\langle \varphi _{M}^{\left( \sigma \right) }\ 
\chi _{M}^{\left(
\sigma \right) }\left| \hat{H}\right| 
\varphi _{M+1}^{\left( \sigma \right)
}\ \chi _{M+1}^{\left( \sigma \right) }\right\rangle =H_{q},
\quad
\left\langle \varphi _{M}^{\left( \sigma \right) }\ 
\chi _{M}^{\left( \sigma
\right) }|\varphi _{M+1}^{\left( \sigma \right) }\ 
\chi _{M+1}^{\left(
\sigma \right) }\right\rangle =S_{q}$;

c) the non-diagonal matrix elements for the states with different 
dipole moment but the same quadrupole one:

$\left\langle \varphi _{M}^{\left( \sigma \right) }\ 
\chi _{M}^{\left(
\sigma \right) }\left| \hat{H}\right| 
\varphi _{M+2}^{\left( \sigma \right)
}\ \chi _{M+2}^{\left( \sigma \right) }\right\rangle =
H_{d},\quad
\left\langle \varphi _{M}^{\left( \sigma \right) }\ 
\chi _{M}^{\left( \sigma
\right) }|\varphi _{M+2}^{\left( \sigma \right) }\ 
\chi _{M+2}^{\left(\sigma \right) }\right\rangle =S_{d}$.

The explicit expressions for these quantities at $V_{z}=0$ are listed 
in Table \ref{table3}. Thus, the system Eq. (\ref{tunnel system}) is written as:
\begin{equation}
\left(
\begin{array}{cccc}
H & H_{q} & H_{d} & H_{q} \\
H_{q} & H & H_{q} & H_{d} \\
H_{d} & H_{q} & H & H_{q} \\
H_{q} & H_{d} & H_{q} & H
\end{array}
\right) \left(
\begin{array}{c}
c_{1} \\
c_{2} \\
c_{3} \\
c_{4}
\end{array}
\right) =E\ \left(
\begin{array}{cccc}
1 & S_{q} & S_{d} & S_{q} \\
S_{q} & 1 & S_{q} & S_{d} \\
S_{d} & S_{q} & 1 & S_{q} \\
S_{q} & S_{d} & S_{q} & 1
\end{array}
\right) \left(
\begin{array}{c}
c_{1} \\
c_{2} \\
c_{3} \\
c_{4}
\end{array}
\right)
\end{equation}
with eigenvectors
\begin{eqnarray}
&& \vec{c}_{1} = \frac{1}{2}\left( 1,1,1,1\right) ,
\quad \vec{c}_{1} = \frac{
1}{2}\left( 1,-1,1,-1\right) ,  \nonumber \\
&& \vec{c}_{3} = \frac{1}{\sqrt{2}}\left( -\sin \theta ,
\cos \theta ,
\sin\theta , -\cos \theta \right) , \\
&& \vec{c}_{4} = \frac{1}{\sqrt{2}}
\left( \cos \theta ,\sin \theta ,
-\cos\theta ,-\sin \theta \right)  \nonumber
\end{eqnarray}
Vectors $\vec{c}_{3}$ and $\vec{c}_{4}$ 
are degenerated, hence there is
a freedom in choice of $\theta $. We assume $\theta =0$, then
\begin{eqnarray}
&& \left| \Psi _{A_{1g}}\right\rangle = c_{\sigma }\left|
A_{1g}\right\rangle \ \chi _{a_{1g}}+\frac{d_{\sigma }}{\sqrt{2}}
\left\{\left| E_{u}^{\left( 1\right) }\right\rangle \ 
\chi _{e_{u}^{\left( 1\right)
}}+\left| E_{u}^{\left( 2\right) }\right\rangle 
\ \chi _{e_{u}^{\left(
2\right) }}\right\} ,  \nonumber \\
&& \left| \Psi _{\Sigma _{{}}^{{}}}\right\rangle = 
-c_{\sigma }\left|
A_{1g}\right\rangle \ \chi _{\Sigma }-
\frac{d_{\sigma }}{\sqrt{2}}\left\{
\left| E_{u}^{\left( 1\right) }\right\rangle \ 
\chi _{e_{u}^{\left( 1\right)
}}-\left| E_{u}^{\left( 2\right) }\right\rangle \ 
\chi _{e_{u}^{\left(
2\right) }}\right\} , \\
&& \left| \Psi _{E_{u}^{\left( 1\right) }}\right\rangle = 
c_{\sigma }\left|
A_{1g}\right\rangle \ \chi _{e_{u}^{\left( 1\right) }}+
\frac{d_{\sigma }}{
\sqrt{2}}\left| E_{u}^{\left( 1\right) }\right\rangle \left\{ \ 
\chi_{a_{1g}}+\chi _{\Sigma }\right\} ,  \nonumber \\
&& \left| \Psi _{E_{u}^{\left( 2\right) }}\right\rangle = 
c_{\sigma }\left|
A_{1g}\right\rangle \ \chi _{e_{u}^{\left( 2\right) }}+
\frac{d_{\sigma }}{
\sqrt{2}}\left| E_{u}^{\left( 2\right) }\right\rangle \left\{ \ 
\chi_{a_{1g}}-\chi _{\Sigma }\right\} ,  \nonumber
\end{eqnarray}
where functions 
$\left\{ \left| E_{u}^{(1)}\right\rangle ,
\left|E_{u}^{(2)}\right\rangle \right\}$ coincide, respectively, with 
$\left\{\left| 00E_{u}^{x}\right\rangle ,
\left| 00E_{u}^{y}\right\rangle \right\}$,
if $\sigma =\alpha $ or with 
$\{(\left| 00E_{u}^{x}\right\rangle +
\left|00E_{u}^{y}\right\rangle )/\sqrt{2}$, 
$(-\left| 00E_{u}^{x}\right\rangle+
\left| 00E_{u}^{y}\right\rangle )/\sqrt{2}\}$, if $\sigma =\beta $.
Symmetric combinations of vibrational functions are:
\begin{eqnarray}
&& \chi _{a_{1g}} = 
\frac{1}{2}\left( \chi _{1}^{\left( \sigma \right)}+
\chi _{2}^{\left( \sigma \right) }+
\chi _{3}^{\left( \sigma \right) }+
\chi_{4}^{\left( \sigma \right) }\right) ,\quad 
\chi _{e_{u}^{\left( 1\right) }}=
\frac{1}{\sqrt{2}}\left( \chi _{2}^{\left( \sigma \right) }-
\chi_{4}^{\left( \sigma \right) }\right) , \\
&& \chi_{\Sigma } = 
\frac{1}{2}\left( -\chi _{1}^{\left( \sigma \right)}+
\chi_{2}^{\left( \sigma \right) }-
\chi _{3}^{\left( \sigma \right) }+
\chi_{4}^{\left( \sigma \right) }\right) ,\quad 
\chi _{e_{u}^{\left( 2\right) }}=
\frac{1}{\sqrt{2}}\left( \chi_{1}^{\left( \sigma \right) }-
\chi_{3}^{\left( \sigma \right) }\right) .  \nonumber
\end{eqnarray}
The symmetry $\Sigma $ of the vibronic and vibrational 
functions coincide
with that of $\sigma $ mode. With taking account of the 
normalization for tunnel states
\begin{eqnarray}
&& N_{A_{1g}}^{2} = 
\left\langle \Psi _{A_{1g}}|\Psi _{A_{1g}}\right\rangle
=1+2S_{q}+S_{d}\ ,  \nonumber \\
&& N_{E_{u}}^{2} = 
\left\langle 
\Psi _{E_{u}^{\left( 1\right) }}|
\Psi_{E_{u}^{\left( 1\right) }}\right\rangle =
1-S_{d}\ , \\
&& N_{\Sigma }^{2} = 
\left\langle \Psi _{\Sigma }|
\Psi _{\Sigma}\right\rangle =1-2S_{q}+S_{d}\ ,  \nonumber
\end{eqnarray}
we come to following expressions for the tunnel energy levels:
\begin{equation}
E_{A_{1g}}=\frac{H+2H_{q}+H_{d}}{1+2S_{q}+S_{d}}\ ,\ 
E_{E_{u}}=\frac{H-H_{d}}{1-S_{d}},\ 
E_{\Sigma }=\frac{H-2H_{q}+H_{d}}{1-2S_{q}+S_{d}}\ .
\end{equation}

These are shown in Fig.\ref{fig6} as a function of $\Delta $ with 
dimensionless coupling constant $k_e=3$. The frequencies of tunneling 
between the equivalent distorted configurations of the $CuO_{4}$ cluster 
are determined by the splittings of the tunnel energy levels, which are 
much less than typical phonon energies. It should be noted, that for 
the pseudo-Jahn-Teller effect the symmetry of the ground vibronic and 
bare electronic states could be different unlike the conventional 
Jahn-Teller effect (''Ham's law'') \cite {Ham1965}. Fig.\ref{fig6} 
illustrates the situation, when $\Delta <0$ (electronic $^{1}A_{1g}$ 
level is higher than $^{1}E_{u}$), but $E_{A_{1g}}<E_{E_{u}}$. The reason 
is that in the pseudo-effect the vibronic interaction mixes different 
electronic levels, contrary to a case of the degenerated electronic 
states.  It is worthy to note that in our case the vibronic $\Sigma$ 
level is always higher in energy than the $A_{1g}$ and $E_{u}$ ones.

\subsection{Degeneracy of JT energies for rhombic modes.}

When $E_{JT}^{\alpha }=E_{JT}^{\beta }$ (and with $a_{\sigma }>0$, 
$b>0$) the trough of minima appears on the lower AP sheet. It is a 
curve in four-dimensional space, which has following parametric form:
\begin{eqnarray}
&& Q_{z} = -q_{z}^{\left( 0\right) }c_{\sigma },\quad 
Q_{x}=-2q_{e}^{\left(0\right) }c_{\sigma }d_{\sigma }
\cos \frac{\varphi }{2},\quad
Q_{y}=-2q_{e}^{\left( 0\right) }c_{\sigma }d_{\sigma }
\sin \frac{\varphi }{2}, \\
&& Q_{\alpha } = -2q_{\sigma }^{\left( 0\right) }d_{\sigma }^{2}
\cos \varphi,\quad 
Q_{\beta }=-2q_{\sigma }^{\left( 0\right) }d_{\sigma }^{2}
\sin\varphi ,  \nonumber
\end{eqnarray}
where $c_{\sigma }=\sqrt{\frac{a_{\sigma }}{a_{\sigma }+b}}$, 
$d_{\sigma }=\sqrt{\frac{b}{a_{\sigma }+b}}$. 
The energy and  wave function are
\begin{eqnarray}
&& \varepsilon _{0}=-E_{JT}^{\sigma }-
\frac{1}{2}c_{\sigma }^{2}a_{\sigma }\,
\\
&& \left| \Phi \right\rangle = 
c_{\sigma }\left| A_{1g}\right\rangle
+d_{\sigma }\left( \cos \frac{\varphi }{2}\left| 
E_{u}^{x}\right\rangle
+\sin \frac{\varphi }{2}\left| 
E_{u}^{y}\right\rangle \right) ,  \nonumber
\end{eqnarray}
respectively.
The $CuO_{4}$ cluster distortions as a function of $\varphi$ 
are shown in Fig.\ref{fig7}.

For JT$_{\sigma}$ case, when $c_{\sigma }=0$, the situation is similar 
to the $E-e$-problem with the doublet ground vibronic state. Its type 
are not changed in PJT$_{\sigma }$ case, if $c_{\sigma }\ll 1$. 
However, with the increasing of $c_{\sigma }$ from the one hand the 
continuum of $E-e$ type splits into two parts in $e_{u}$ direction and 
from the other hand the $A_{1g}-E_{u}$ mixing reduces the symmetry of 
electronic wave function resulting in a singlet ground state. The 
potential energy in a small vicinity of the continuum minima is
\begin{equation}
\varepsilon =
\varepsilon _{0}+\frac{\omega _{z}^{2}q_{z}^{2}}{2}+
\frac{Dr^{2}}{2}+\frac{E\rho ^{2}}{2}-Fr\rho \ ,  \label{radial QF}
\end{equation}
where $q_{z}$, $r$ and $\rho $ are count out from
\[
Q_{z}^{\left( 0\right) }=-q_{z}^{\left( 0\right) }c_{\sigma }^{2}\ ,
\quad
r_{0}=-2q_{e}^{\left( 0\right) }c_{\sigma }d_{\sigma }\ ,\quad 
\rho_{0}=-q_{\sigma }^{\left( 0\right) }d_{\sigma }^{2}\ .
\]

The radial vibration frequencies along the principal directions of the
quadratic form Eq. (\ref{radial QF}) are $\omega _{+}^{\prime }$ and 
$\omega _{-}^{\prime }$ Eq. (\ref{freq2}), respectively. If 
$r_{0}^{2}+\rho _{0}^{2}\gg \hbar /\omega _{-}^{\prime }$, 
$\varepsilon _{0}\gg \hbar \omega _{+}^{\prime }$, then the energy and 
wave function of the ground vibronic singlet are
\begin{equation}
E_{0}=\varepsilon_{0}+
\frac{\hbar }{2}\left( \omega _{+}^{\prime }+
\omega_{-}^{\prime }\right) \ ,\quad 
\left| \Psi _{0}\right\rangle =\left| \Phi
\right\rangle \frac{\chi _{0}\left( r_{+}\right) \ 
\chi _{0}\left( \rho_{-}\right) }{\sqrt{2\pi 
\left( r_{0}+r\right) 
\left( \rho _{0}+\rho \right)}}\ ,  \label{rot_singlet}
\end{equation}
where $\chi _{0}$ is the ground state vibrational function.

If the parameters approach to those typical for the NJT situation, the
radius of trough is small, so the rotational term is not a small
perturbation. The energy barrier in the higher symmetry point lowers, 
and the system turns into vibrational regime near the NJT minima with 
a singlet ground state, which is not described by the Eq. 
(\ref{rot_singlet}).

\section{Vibronic states in  presence of the spin-orbit coupling.}

Without taking account of the spin-orbit coupling the $^3E_u$ term is
isolated, and it has the six-fold degenerated ground vibronic state. 
The spin-orbit coupling mixes the $M_S=0$ states of the $^3E_u$ and 
$^1E_u$ terms. As well, this coupling splits lower vibronic states of 
the $^3E_u$ term, which have $M_S=\pm 1$.

\subsection{ The $M_S=0$ states.}

\subsubsection{Well-isolated $^{1,3}E_u$ terms.}

If the $^{1}A_{1g}$ and $^{1,3}E_{u}$ terms are well separated in 
energy ($\Delta _{AE}\gg \Delta _{E},\lambda $), it is possible to 
consider the AP within a basis of the 
$\left| 00E_{u}^{x}\right\rangle$, 
$\left| 00E_{u}^{y}\right\rangle $, 
$\left| 10E_{u}^{x}\right\rangle $, 
$\left| 10E_{u}^{y}\right\rangle $ states, 
and then take into account the vibronic coupling with the of 
$^{1}A_{1g}$ term as perturbation. The potential energy matrix 
$\hat{U}(Q)$ acquires a form:

\begin{equation}
\sum_{i}\frac{\omega _{i}^{2}Q_{i}^{2}}{2}\cdot \hat{I}+\left(
\begin{array}{cccc}
-\Delta _{E}+V_{\alpha }Q_{\alpha } & 
V_{\beta }Q_{\beta } & 0 & -i\lambda \\
V_{\beta }Q_{\beta } & 
-\Delta _{E}-V_{\alpha }Q_{\alpha } & i\lambda & 0 \\
0 & -i\lambda & 
\Delta _{E}+V_{\alpha }Q_{\alpha } & V_{\beta }Q_{\beta } \\
i\lambda & 0 & 
V_{\beta }Q_{\beta } & \Delta _{E}-V_{\alpha }Q_{\alpha }
\end{array}
\right) ,
\end{equation}
where $\lambda $ is  submatrix element of the spin-orbit coupling. The
eigenvalues and eigenvectors of $\hat{U}(Q)$ are written as follows
\begin{equation}
\begin{array}{ll}
\varepsilon _{1}=\Sigma -
\sqrt{\left( \Delta _{E}+\rho \right) ^{2}+\lambda
^{2}},\ \, & \left| 1\right\rangle =-i\sin \eta _{1}\left| -\rho
,0\right\rangle +\cos \eta _{1}\left| \rho ,1\right\rangle \ , \\
\varepsilon _{2}=
\Sigma -\sqrt{\left( \Delta _{E}-\rho \right) ^{2}+\lambda
^{2}}\,,\  & \left| 2\right\rangle =i\sin \eta _{2}\left| \rho
,0\right\rangle +\cos \eta _{2}\left| -\rho ,1\right\rangle \ , \\
\varepsilon _{3}=\Sigma +
\sqrt{\left( \Delta _{E}-\rho \right) ^{2}+\lambda
^{2}}\,,\  & \left| 3\right\rangle =\cos \eta _{2}\left| \rho
,0\right\rangle +i\sin \eta _{2}\left| -\rho ,1\right\rangle \ , \\
\varepsilon _{4}=
\Sigma +\sqrt{\left( \Delta _{E}+\rho \right) ^{2}+\lambda
^{2}}\,,\  & \left| 4\right\rangle =\cos \eta _{1}\left| -\rho
,0\right\rangle -i\sin \eta _{1}\left| \rho ,1\right\rangle \ ,
\end{array}
\end{equation}
where
\begin{eqnarray}
&& \Sigma = 
\sum_{i}\frac{\omega _{i}^{2}Q_{i}^{2}}{2},\quad \rho =\sqrt{
\left( V_{\alpha }Q_{\alpha }\right)^{2}+
\left( V_{\beta }Q_{\beta }\right)^{2}}, \\
&& \left| \rho ,S\right\rangle = 
\cos \theta \left| S0E_{u}^{x}\right\rangle
+\sin \theta \left| S0E_{u}^{y}\right\rangle ,\   \nonumber \\
&& \left| -\rho ,S\right\rangle = -\sin \theta \left|
S0E_{u}^{x}\right\rangle +
\cos \theta \left| S0E_{u}^{y}\right\rangle ,
\nonumber \\
&& \tan 2\theta = 
\frac{V_{\beta }Q_{\beta }}{V_{\alpha }Q_{\alpha }}\,\quad 
\tan 2\eta _{1}=
\frac{\lambda }{-\Delta _{E}-\rho }\ ,\quad \tan
2\eta _{2}=\frac{\lambda }{-\Delta _{E}+\rho }\ .  \nonumber
\end{eqnarray}

The minima of the lower AP sheet  are located on $Q_{\sigma }$ axis
(with $E_{JT}^{\sigma }>E_{JT}^{\sigma ^{\prime }}$, $\sigma ,\sigma
^{\prime }=\alpha ,\beta $) at points, which represent solutions of the
equation:
\begin{equation}
\frac{\left| Q_{\sigma }\right| }{q_{\sigma }^{\left( 0\right) }}=
\left(1+\frac{\lambda^2}{\left( V_{\sigma }\left| Q_{\sigma }\right| +
\Delta_{E}\right) ^{2}}\right) ^{-\frac{1}{2}}.
\end{equation}

The $Q_{\Gamma \gamma }=0$ is a point of discontinuity of the 
derivative.  The non-trivial minima, which correspond to the 
low-symmetry cluster distortions, exist for the arbitrary large 
$\lambda $, if $\Delta _{E}\neq 0$,  but with $\lambda \rightarrow 
\infty $ the minima depth and the distortions magnitude become 
negligibly small.  The ground vibronic state is twice degenerated. In 
strong coupling scheme this is realized due to orthogonality of 
electronic states belonging to different minima of AP. In a case 
$\lambda \ll E_{JT}^\sigma $ the expressions for the minimum points 
and their energy are written as follows:
\begin{eqnarray}
&& Q_{\sigma } = \pm \tilde{q}_{\sigma }^{\left( 0\right) },\quad 
\tilde{q}_{\sigma }^{\left( 0\right) }=
q_{\sigma }^{\left( 0\right) }
\left( 1-\frac{\lambda ^{2}}{2\left( \Delta _{E}+
2E_{JT}^{\sigma }\right) ^{2}}\right) , \\
&& \varepsilon _{0} = -\Delta _{E}-E_{JT}^{\sigma }-
\frac{\lambda^{2}}{2\left( \Delta _{E}+
2E_{JT}^{\sigma }\right) }\ .  \nonumber
\end{eqnarray}
The mixing coefficient for the triplet spin states 
with the singlet $^{1}A_{1g}$
term wave function is proportional to
\begin{equation}
\frac{\lambda }{\Delta _{E}+\Delta _{AE}+E_{JT}^{e}}\ .
\end{equation}

\subsubsection{Strong PJT-effect for the singlet spin states.}

In this case it is neccessary to consider the spin-orbit mixing of the
tunnel states with lower vibronic states of $^3E_u$ term. The only 
non-zero matrix elements are those between the tunnel $E_u$-states and 
the lower vibronic states of $^3E_u$ term with $M_S=0$. The effective 
Hamiltonian matrix is

\begin{equation}
\left(
\begin{array}{cccc}
-\Delta _{E}^{\prime } & 0 & 0 & -i\lambda ^{\prime } \\
0 & -\Delta _{E}^{\prime } & i\lambda ^{\prime } & 0 \\
0 & -i\lambda ^{\prime } & \Delta _{E}^{\prime } & 0 \\
i\lambda ^{\prime } & 0 & 0 & \Delta _{E}^{\prime }
\end{array}
\right) ,
\end{equation}
where $2\Delta _{E}^{\prime }$ is an appropriate 
energy splitting, $\lambda^{\prime }$ is a 
modified matrix element of the spin-orbit coupling:
\begin{equation}
\lambda ^{\prime }=\ 
\lambda \ N_{E_{u}}^{-1}\sqrt{2}d_{\sigma }\left\langle
\chi _{M}^{\left( \sigma \right) }|
\tilde{\chi}_{0}^{\left( \sigma \right)}\right\rangle \ ,
\end{equation}
where $\chi _{M}^{\left( \sigma \right) }$ and 
$\tilde{\chi}_{0}^{\left( \sigma \right) }$ are the ground state 
vibrational functions, which are centered at the minimum $M$ of 
 PJT$_{\sigma }$ type and at the minimum of AP of the 
 $E-b_{1}-b_{2}$-problem with opposite to $M$ sign of $\sigma $ mode, 
 respectively.

\subsubsection{The singlet-triplet asymmetry 
of the vibronic coupling.}

Let the difference in the linear vibronic coupling constants for the 
singlet $^1E_u$ and triplet $^3E_u$ states corresponds to the 
following inequalities for the JT energies: $E_{JT}^\alpha 
(^1E_u)>E_{JT}^\beta (^3E_u)$ and $E_{JT}^\alpha (^1E_u)<E_{JT}^\beta 
(^3E_u)$. Then the minima of  AP for the $^1E_u$ term are located on 
the $Q_\alpha $ axis, and those of the $^3E_u $ therm are located on 
the $Q_\beta $ axis. If the surfaces of  AP for different terms 
intersect, than the taking account of the spin-orbit coupling could 
result in a complicated form of the AP with four minima. The states in 
minima located on the same axis are orthogonal to each other, and 
those located on different axis are not. Hence, even if $\Delta _E=0$ 
the lower vibronic states are two doublets, which are separated by the 
tunnel splitting $\tilde{\Delta }$. The frequency related to 
$\tilde{\Delta } $ corresponds to the combined pulsing motion of the 
electronic and nuclear density between $b_{1g}$ and $b_{2g}$ 
distortions of the $CuO_4$ cluster.

\subsection{The $M_{S}=\pm 1$ states.}

A joint operation of the vibronic and spin-orbital coupling for the 
$M_{S}=\pm 1$ states within the $^{3}E_{u}$ term is described 
by the matrix
\begin{equation}
V_{\alpha }Q_{\alpha }\hat{\sigma}_{z}+V_{\beta }
Q_{\beta }\hat{\sigma}_{x}+iM_{S}\lambda _{1}\hat{\sigma}_{y}\ ,
\end{equation}
where $\lambda _{1}$ is a submatrix element for the spin-orbital 
coupling within the orbital part of the $^{3}E_{u}$ manifold, 
$\hat{\sigma}_{i}$ are the Pauli matrices, and the energy is counted 
off the $^{3}E_{u}$ manifold.

The lower AP sheet has four extrema at points
\begin{equation}
q_{i}^{\left( 0,\pm \right) }=
\pm l_{i}\sqrt{2\tilde{E}_{JT}^{i}\left(
1-p_{i}^{2}\right) }\ ,\quad i=\alpha ,\beta ,
\end{equation}
where $l_{i}=\sqrt{\hbar /\omega _{i}}$, 
$\tilde{E}_{JT}^{i}=E_{JT}^{i}/\hbar \omega _{i}$, 
$p_{i}=\lambda _{1}/2E_{JT}^{i}$, and 
$q_{i}^{\left(0,\pm \right) }=0$ at 
$\left| p_{i}\right| \geq 1$. The parameter $p_{i}$
equals to the ratio of the $E_{u}$ level splitting due to the 
spin-orbital coupling ($2\lambda _{1}$) 
to that of due to the vibronic coupling ($4E_{JT}^{i}$ ). At 
$\left| p_{i}\right| \geq 1$ ($i=\alpha ,\beta $) we come
to the weak pseudo-Jahn-Teller effect with the only minimum at 
$q_{i}^{\left( 0,\pm \right) }=0$ ($i=\alpha ,\beta $) and effective
local vibration frequencies
\begin{equation}
\tilde{\omega}_{i}^{2}=\omega _{i}^{2}\left( 1-\left| p_{i}\right|
^{-1}\right) .
\end{equation}

At $\left| p_{i}\right| \leq 1$ ( 
$E_{JT}^{\sigma }>E_{JT}^{\sigma ^{\prime}}$) the 
strong pseudo-Jahn-Teller effect occurs with the AP minima at 
$q_{\sigma }^{\left( 0,\pm \right) }$, and with the AP saddle points at 
$q_{\sigma ^{\prime }}^{\left( 0,\pm \right) }$. 
The effective frequencies for the local vibrations 
at the AP minima and corresponding energies are
derived as follows:
\begin{eqnarray}
&& \tilde{\omega}_{\sigma }^{2} = 
\omega _{\sigma }^{2}\left( 1-p_{\sigma
}^{2}\right) ,\ 
\tilde{\omega}_{\sigma ^{\prime }}^{2}=\omega _{\sigma
^{\prime }}^{2}\left( 1-\lambda _{\sigma }\right) ,\ 
\lambda _{\sigma }=
\frac{E_{JT}^{\sigma ^{\prime }}}{E_{JT}^{\sigma }}\ , \\
&& \varepsilon_{-}
\left( q_{\sigma }^{\left( 0,\pm \right) }\right) =
-E_{JT}^{\sigma }-\hbar \omega _{\sigma }p_{\sigma }\ .  \nonumber
\end{eqnarray}
As is seen, unlike the usual $E-b_{1}-b_{2}$-problem, 
the $\sigma$ mode frequency is renormalized 
due to the spin-orbital coupling.

The electronic wave functions 
$\varphi _{\pm }^{\left( M_{S}\right) }$ at the minima
 $q_{\sigma }^{\left( 0,\pm \right) }$ are reduced to the
following form:
\begin{eqnarray}
\varphi _{+}^{\left( M_{S}\right) }
 &=&\frac{1}{\sqrt{2}}\left( i\ M_{S}
\sqrt{1-\sqrt{1-p_{\sigma }^{2}}}\left| E_{u}^{\left( 1\right)
}\right\rangle +
\sqrt{1+\sqrt{1-p_{\sigma }^{2}}}\left| E_{u}^{\left(
2\right) }\right\rangle \right) , \\
\varphi _{-}^{\left( M_{S}\right) }
 &=&\frac{1}{\sqrt{2}}\left( \sqrt{1+
\sqrt{1-p_{\sigma }^{2}}}\left| 
E_{u}^{\left( 1\right) }\right\rangle -i\
M_{S}\sqrt{1-\sqrt{1-p_{\sigma }^{2}}}\left| 
E_{u}^{\left( 2\right)
}\right\rangle \right) .  \nonumber
\end{eqnarray}
In contrast to the above considered 
$E-b_{1}-b_{2}$-problem,  the spin-orbital coupling now 
results in the non-orthogonality of the wave
functions located in  the minima:
\begin{equation}
\left\langle \varphi _{+}^{\left( M_{S}\right) }|\varphi _{-}^{\left(
M_{S}\right) }\right\rangle =-i\ M_{S}\ p_{\sigma }
\end{equation}
that leads to the tunnel splitting.

Varying the energy functional with the basis functions
\begin{equation}
\Psi =a\ \varphi _{+}^{\left( M_{S}\right) }\chi _{+}+b\ \varphi
_{-}^{\left( M_{S}\right) }\chi _{-},
\end{equation}
where $\chi _{\pm }$ are the vibrational functions centered at 
$q_{\sigma}^{\left( 0,\pm \right) }$, results in a matrix equation:
\begin{equation}
\left(
\begin{array}{cc}
H_{++} & H_{+-} \\
H_{+-}^{*} & H_{++}
\end{array}
\right) \left(
\begin{array}{c}
c_{1} \\
c_{2}
\end{array}
\right) =E\ \left(
\begin{array}{cc}
1 & S_{+-} \\
S_{+-}^{*} & 1
\end{array}
\right) \left(
\begin{array}{c}
c_{1} \\
c_{2}
\end{array}
\right) ,
\end{equation}
where
\begin{eqnarray}
&& S = \left\langle 
\varphi _{+}^{\left( M_{S}\right) }\chi _{+}|
\varphi_{-}^{\left( M_{S}\right) }\chi _{-}\right\rangle =
-i\ M_{S}\ p_{\sigma }\
\exp \left[ -2\tilde{E}_{JT}^{\sigma }
\left( 1-p_{\sigma }^{2}\right) ^{
\frac{3}{2}}\right],  \label{tunnel matix element} \\
&& H_{++} = \hbar \omega _{\sigma }\left[ 
\frac{2-p_{\sigma }^{2}}{4\sqrt{1-p_{\sigma }^{2}}}-
\tilde{E}_{JT}^{\sigma }\left( 1+p_{\sigma }^{2}\right)
\right] +\hbar \omega_{\sigma ^{\prime }}
\frac{2-\lambda _{\sigma }}{4\sqrt{
1-\lambda _{\sigma }}}\ ,  \nonumber \\
&& H_{+-} = \left( H_{++}-\hbar 
\omega _{\sigma }B\right) S\ ,\quad 
B=\tilde{E}_{JT}^{\sigma }\left( 1-p_{\sigma }^{2}\right) 
\left( 2-p_{\sigma}^{2}\right) .  \nonumber
\end{eqnarray}

The tunnel level energies and tunnel splitting are determined 
as follows:
\begin{eqnarray}
&& E_{g} = 
H_{++}-\frac{\hbar \omega _{\sigma }B\left| S\right| \ }{1+\left|
S\right| }\ ,\quad E_{u}=H_{++}+\frac{\hbar \omega _{\sigma }B\left|
S\right| \ }{1-\left| S\right| }\ , \\
&& \Delta _{1} = E_{g}-E_{u}=
2\hbar \omega _{\sigma }\frac{B\left| S\right|
\ }{1-\left| S\right| ^{2}}\ .  \nonumber
\end{eqnarray}
Appropriate vibronic wave functions can be represented as:
\begin{eqnarray}
&& \Psi _{g}^{\left( 1\right) } = 
N_{g}\left( \varphi _{+}^{\left( 1\right)}\chi_{+}+
i\varphi _{-}^{\left( 1\right) }\chi _{-}\right) , \quad
\Psi_{u}^{\left( 1\right) }=
N_{u}\left( i\varphi _{+}^{\left( 1\right) }\chi_{+}+
\varphi _{-}^{\left( 1\right) }
\chi_{-}\right) ,  \label{tunnel states}
\\
&& \Psi _{g}^{\left( -1\right) } = 
N_{g}\left( i\varphi _{+}^{\left(-1\right) }\chi _{+}+
\varphi _{-}^{\left( -1\right) }\chi _{-}\right) , \,
\Psi_{u}^{\left( -1\right) }=
N_{u}\left( \varphi _{+}^{\left( -1\right)}\chi _{+}+
i\varphi _{-}^{\left( -1\right) }\chi _{-}\right) ,  \nonumber
\end{eqnarray}
where
\[
N_{g}=\frac{1}{\sqrt{2\left( 1+\left| S\right| \right) }},\quad 
N_{u}=\frac{1}{\sqrt{2\left( 1-\left| S\right| \right) }}.
\]
In the limit $\left| p_{\sigma }\right| \ll 1$ for the magnitude of
tunnel splitting one obtains:
\begin{equation}
\Delta_{1} = 
2 \lambda_{1} \exp \left( -2 \tilde{E}_{JT}^{\sigma} \right),
\end{equation}
that might be interpreted as a result of vibronic reduction for the
purely spin-orbital splitting.

\section{Reduction factors.}

A concept of vibronic reduction is widely used in a theory of the 
Jahn-Teller effect. The reduction factor is equal to a ratio of the 
reduced matrix element of the electronic operator, which is calculated 
for vibronic ground states, to this one for the bare electronic 
states. In a case of the pseudo-Jahn-Teller effect the vibronic 
reduced matrix element is commonly a linear combination of the 
electronic ones.

Within the strong coupling scheme the vibronic state is written as
\begin{equation}
\Psi _{\Gamma \gamma }=
\frac{1}{N_{\Gamma }}\sum_{\Gamma _{1}}c\left( \Gamma
\Gamma _{1}\right) 
\sum_{\gamma _{1}\Gamma _{2}\gamma _{2}}\varphi_{\Gamma
_{1}\gamma _{1}}
\chi _{\Gamma _{2}\gamma _{2}}\left\langle \Gamma_{1}\gamma
_{1}\Gamma _{2}\gamma _{2}|\Gamma \gamma \right\rangle \ ,
\end{equation}
where $\Gamma _{2}\in \Gamma \times \Gamma _{1}$, $N_{\Gamma }$ is the
normalization factor; $c\left( \Gamma \Gamma _{1}\right) $ is the 
amplitude of  contribution to the vibronic function $\Gamma $, which 
is generated by the electronic $\Gamma _{1}$ state; $\chi _{\Gamma 
_{2}\gamma _{2}}$ is a symmetric linear combination of the vibrational 
states of equivalent minima; $\left\langle \Gamma _{1}\gamma 
_{1}\Gamma _{2}\gamma _{2}|\Gamma \gamma \right\rangle $ is the 
Clebsh-Gordan coefficient.

Consider the vibronic matrix element of electronic operator
$\hat{V}_{\tilde{\Gamma }\tilde{\gamma }}$
\begin{eqnarray}
&& \left\langle \Psi _{\Gamma \gamma }\left| 
\hat{V}_{\tilde{\Gamma }\tilde{\gamma }}\right| 
\Psi _{\Gamma ^{\prime }\gamma ^{\prime }}\right\rangle =
\frac{1}{N_{\Gamma }N_{\Gamma ^{\prime }}}
\sum_{\Gamma _{1}\Gamma_{1}^{\prime }}
c(\Gamma \Gamma _{1})\ 
c(\Gamma ^{\prime }\Gamma _{1}^{\prime})
\frac{\left\langle \varphi _{\Gamma _{1}}\left\| 
\hat{V}_{\tilde{\Gamma }}\right\| 
\varphi _{\Gamma _{1}}\right\rangle }{\sqrt{g_{\Gamma _{1}}}}\times
\\
&& \sum_{\Gamma _{2}}\left\langle \chi_{\Gamma_{2}}^{2}\right\rangle
\sum_{\gamma _{1}\gamma _{1}^{\prime }\gamma _{2}}\left\langle 
\Gamma_{1}\gamma _{1}|
\tilde{\Gamma }\tilde{\gamma }\Gamma _{1}^{\prime }
\gamma_{1}^{\prime }\right\rangle \left\langle \Gamma_{1}
\gamma _{1}\Gamma_{2}\gamma _{2}|
\Gamma \gamma \right\rangle \left\langle 
\Gamma _{1}^{\prime}\gamma_{1}^{\prime }\Gamma _{2}\gamma _{2}|
\Gamma ^{\prime }\gamma^{\prime }\right\rangle .  \nonumber
\end{eqnarray}

Here we used the Wigner-Eckart-Koster \cite{Koster1958} theorem for 
the electronic matrix element in a case of the simply reducible group:
\begin{equation}
\left\langle \Gamma \gamma \left| 
\tilde{\Gamma }\tilde{\gamma }\right|
\Gamma ^{\prime }\gamma ^{\prime }\right\rangle =
\frac{\left\langle \Gamma
\left\| \tilde{\Gamma }\right\| 
\Gamma ^{\prime }\right\rangle } {\sqrt{
g_{\Gamma }}}\left\langle \Gamma \gamma |
\tilde{\Gamma } \tilde{\gamma}
\Gamma^{\prime } \gamma^{\prime } \right\rangle \ ,
\end{equation}
and the orthogonality relation for the vibrational functions:
\begin{equation}
\left\langle \chi _{\Gamma _{2}\gamma _{2}}|
\chi _{\Gamma _{2}^{\prime}\gamma _{2}^{\prime }}\right\rangle =
\left\langle \chi _{\Gamma_{2}}^{2}\right\rangle \ 
\delta _{\Gamma _{2}\Gamma _{2}^{\prime }}
\delta_{\gamma_{2}\gamma_{2}^{\prime }}\ ,
\end{equation}
where $\left\langle \chi _{\Gamma _{2}}^{2}\right\rangle $ is the
normalization factor. Applying the Wigner-Eckart-Koster theorem for 
the vibronic matrix element we find for real representations:
\begin{equation}
\left\langle \Psi _{\Gamma }\left\| 
\hat{V}_{\tilde{\Gamma }}\right\| 
\Psi _{\Gamma ^{\prime }}\right\rangle =\sum_{\Gamma _{1}\Gamma
_{1}^{\prime }}K_{\tilde{\Gamma }}\left(
\begin{array}{cc}
\Gamma & \Gamma ^{\prime } \\
\Gamma _{1} & \Gamma _{1}^{\prime }
\end{array}
\right) \left\langle \varphi _{\Gamma _{1}}\left\| 
\hat{V}_{\tilde{\Gamma }}\right\| 
\varphi _{\Gamma _{1}^{\prime }}\right\rangle \ ,
\end{equation}
where
\begin{eqnarray}
K_{\tilde{\Gamma }}\left(
\begin{array}{cc}
\Gamma & \Gamma ^{\prime } \\
\Gamma _{1} & \Gamma _{1}^{\prime }
\end{array}
\right) &=&\frac{\sqrt{g_{\Gamma }
g_{\Gamma ^{\prime }}}}{N_{\Gamma}N_{\Gamma ^{\prime }}}
(-1)^{\Gamma +\tilde{\Gamma }+\Gamma_{1}}
c(\Gamma \Gamma _{1})\ c(\Gamma ^{\prime }\Gamma _{1}^{\prime }) \times
\label{RedFac} \\
&&\sum_{\Gamma _{2}}(-1)^{\Gamma _{2}}\left\langle \chi _{\Gamma
_{2}}^{2}\right\rangle \left[
\begin{array}{ccc}
\Gamma & \tilde{\Gamma } & \Gamma ^{\prime } \\
\Gamma _{1}^{\prime } & \Gamma _{2} & \Gamma _{1}
\end{array}
\right] \ .  \nonumber
\end{eqnarray}
Here $\left[\right]$ is the 6$\Gamma$-symbol for point group G 
\cite{Sviridov1964}.  The factor $(-1)^{\Gamma }$ is 1 for all 
irreducible represenation of $D_{4h} $ except $(-1)^{A_{2g}}=-1$ 
\cite{Hamermesh1964}. In the usual Jahn-Teller situation Eq. 
(\ref{RedFac}) gives the well-known result for the vibronic reduction 
factors \cite{Englman1970}. Note that
\begin{equation}
K_{\tilde{\Gamma }}\left(
\begin{array}{cc}
\Gamma & \Gamma ^{\prime } \\
\Gamma _{1} & \Gamma _{1}^{\prime }
\end{array}
\right) =(-1)^{\Gamma +\Gamma ^{\prime }+
\Gamma _{1}+\Gamma _{1}^{\prime }}\
K_{\tilde{\Gamma }}\left(
\begin{array}{cc}
\Gamma ^{\prime } & \Gamma \\
\Gamma _{1}^{\prime } & \Gamma _{1}
\end{array}
\right) .  \label{symRF}
\end{equation}

In our problem $\Gamma _{1},\Gamma _{1}^{\prime }=A_{1g},E_{u}$; 
$\Gamma ,\Gamma ^{\prime },\Gamma _{2},\Gamma _{2}^{\prime 
}=A_{1g},\Sigma ,E_{u}$; only the operators of symmetry 
$\tilde{\Gamma} = 
A_{1g},A_{2g},B_{1g},B_{2g},E_{u}$ have the non-zero matrix 
elements. The non-trivial reduction factors are given in Table \ref{table4}; 
the remaining ones could be derived with help of Eq. (\ref{symRF}). The 
reduction factors as functions of $\Delta $ are shown in Fig.\ref{fig8}a. 
One can see that $\Sigma ^{\prime}$ operators are reduced very strongly. 
With an increase in the vibronic coupling the reduction factors tend 
to their limit values (Fig.\ref{fig8}b), which can be obtained from Eq. 
(\ref{RedFac}) with $N_{\Gamma }\rightarrow 1$, 
$\left\langle \chi_{\Gamma _{2}}^{2}\right\rangle \rightarrow 1$.

The components of external electric field parallel to the $CuO_4$
 cluster plane ($\vec E \perp \,C_4$) induce an electro-dipole 
transitions between tunnel $A_{1g}$ and $E_u$ states. The reduced 
vibronic matrix element of the $E_u$ operator is written as:
\begin{equation}
\left\langle \Psi _{A_{1g}}\left\| \hat{V}_{E_u}\right\| \Psi
_{E_u}\right\rangle =\frac{c_\sigma d_\sigma }{N_{A_{1g}}N_{E_u}}\left(
\left\langle \chi _{A_{1g}}^2\right\rangle +\left\langle \chi
_{E_u}^2\right\rangle \right) \left\langle A_{1g}\left\| \hat{V}
_{E_u}\right\| E_u\right\rangle .
\end{equation}
Thus, at strong vibronic coupling ($N_\Gamma \rightarrow 1$, 
$\left\langle \chi _{\Gamma _2}^2\right\rangle \rightarrow 1$) and at 
$c_\sigma =d_\sigma =1/\sqrt{2}$ the matrix element of electro-dipole 
transition is not renormalized.

For the singlet $E_{u}$ states an orbital contribution to the Zeeman 
energy is not zero only for the non-zero values of the $z$-component 
of  external magnetic field. This situation could be examined with a 
help of an effective spin Hamiltonian for the non-Kramers doublet 
\cite{Abragam}. The vibronic coupling results in an essential 
renormalization of reduced matrix element of the $z$-component of 
angular momentum:
\begin{equation}
\left\langle \Psi _{E_{u}}\left\| \hat{V}_{A_{2g}}\right\| 
\Psi_{E_{u}}\right\rangle =
\frac{\sqrt{2}d_{\sigma }^{2}S_{q}^{n}}{N_{E_{u}}^{2}}
\left\langle E_{u}\left\| \hat{V}_{A_{2g}}\right\| 
E_{u}\right\rangle .
\end{equation}
In this case the vibronic reduction factor is proportional to overlap
integral $S_{q}^{n}$ of the vibrational states centered at neighboring
minima on the lower AP sheet. With the increasing in the vibronic 
coupling $S_{q}^{n}\rightarrow 0$, that leads to a complete quenching 
of the angular momentum.

\section{Spin Hamiltonian of the PJT center.}

A detailed pattern of the energy spectrum in  external magnetic field 
 and magneto-resonance properties of the PJT centers are substantially 
depend on the bare $A-E$ splitting $\Delta $, vibronic parameters and 
relative magnitude of the vibronic and spin-orbital coupling.

However, some common features are determined only by the symmetry
considerations, in particular, by the $C_4$ axial symmetry and  
specific properties of the bare electronic basis functions. So, the 
orbital doublet $^{1,3}E_u$ terms are characterized by the nonquenched 
highly anisotropic (Izing like) effective orbital moment 
$\tilde{l}=1/2$ oriented only along the $C_4$ axis.

Without taking account of vibronic coupling the effective spin 
Hamiltonian for the $^{1,3}E_{u}$ terms could be represented as 
follows:
\begin{equation}
H=\lambda _{1}\tilde{l}_{z}\hat{S}_{z}+
\beta H_{z}g_{l}\tilde{l}_{z}\hat{I}+2\beta 
\sum_{i=x,y,z}H_{i}\hat{S}_{i},
\end{equation}
where $\beta $ is the Bohr magneton, $\hat{S}_{i}$ are spin matrices 
($S=0$ for the spin singlet $^{1}E_{u}$ term), $g_{l}$ is an effective 
orbital $g$ -factor for the $^{1,3}E_{u}$ term, which magnitude is 
determined by the structure of  electronic $e_{u}$ function. Here, the 
first term describes the spin-orbital coupling, the second and the 
third ones correspond to the orbital and spin Zeeman coupling, 
respectively, with the purely isotropic spin $g$-tensor: 
$g_{x}=g_{y}=g_{z}=2$. Taking account of the spin-orbital 
$^{1}E_{u}$-$^{3}E_{u}$ mixing leads to  emergence of the spin 
anisotropy with  additive contribution to the effective spin 
Hamiltonian:

\begin{equation}
V_{an}=D\hat{S}_{z}^{2}\ ,\quad D=\Delta _{st}-\sqrt{\Delta
_{st}^{2}+\lambda ^{2}}\approx -\frac{\lambda ^{2}}{2\Delta _{st}}\ ,
\end{equation}
and to the effective axial anisotropy of the spin $g$-tensor: 
$g_{z}=2$, $g_{x}=g_{y}=2\cos \theta $, where
\begin{equation}
\cos \theta =
\frac{1}{\sqrt{2}}\sqrt{1+
\frac{\Delta _{st}}{\sqrt{\Delta_{st}^{2}+
\lambda ^{2}}}}\approx 1-\frac{\lambda ^{2}}{8\Delta _{st}^{2}}\ .
\end{equation}

Taking account of the vibronic coupling upon the conditions of  weak
pseudo-Jahn-Teller effect ($p_{\sigma }=\lambda _{1}/2E_{JT}^{\sigma 
}>1$) does not vary a form of the effective spin Hamiltonian since the 
vibronic distortions are suppressed by the spin-orbital coupling. With 
the strong pseudo-Jahn-Teller effect ( $p_{\sigma }=\lambda 
_{1}/2E_{JT}^{\sigma }<1$) one should make use of the vibronic states 
(\ref{tunnel states}), that results in a very complicated 
spin-vibronic effective Hamiltonian. A relatively simple situation 
occurs at $p_{\sigma }^{2}\ll 1$, and small magnitude of the overlap 
for the vibrational functions, when, neglecting the spin-orbital 
$^{1}E_{u}$-$^{3}E_{u}$ mixing, we come to an effective Hamiltonian:
\begin{equation}
H=-\beta H_{z}g_{1}\left\langle 
\chi _{+}|\chi _{-}\right\rangle \tilde{l}_{z}\hat{I}+
\beta \sum_{i=x,y,z}H_{i}g_{ij}\hat{S}_{j}\ ,
\end{equation}
where the overlap integral 
$\left\langle \chi _{+}|\chi _{-}\right\rangle$ 
is determined according to Eq. (\ref{tunnel matix element}) 
and for the axial $g$-tensor: $g_{z}=2-p_{\sigma }g_{1}$, 
$g_{x}=g_{y}=2$. It should be noted the principal difference in the 
$g$-factor anisotropy with and without vibronic effects.

The doublet of the states with different ($\pm $) projections of the
effective orbital moment corresponds to the functions 
\{$\Psi _{g}^{\left(1\right) }$, 
$\Psi _{g}^{\left( 0\right) }$, 
$\Psi _{u}^{\left( -1\right) }$\} and 
\{$\Psi _{u}^{\left( 1\right) }$, 
$\Psi _{u}^{\left( 0\right) }$, 
$\Psi _{g}^{\left( -1\right) }$\}, 
respectively, where, along with the above
defined functions Eq. (\ref{tunnel states}), we have introduced

\begin{eqnarray}
\Psi _{g}^{\left( 0\right) } &=&\frac{1}{\sqrt{2}}\left( \left|
E_{u}^{\left( 2\right) }\right\rangle \ 
\chi _{+}^{\left( 0\right) }+i\left|
E_{u}^{\left( 1\right) }\right\rangle \ 
\chi _{-}^{\left( 0\right) }\right) ,
\\
\Psi _{u}^{\left( 0\right) } &=&\frac{1}{\sqrt{2}}\left( i\left|
E_{u}^{\left( 2\right) }\right\rangle \ 
\chi _{+}^{\left( 0\right) }+\left|
E_{u}^{\left( 1\right) }\right\rangle \ 
\chi _{-}^{\left( 0\right) }\right) .
\nonumber
\end{eqnarray}
An upper label for the vibrational function underlines the difference 
in the location of the minima for the $M_{S}=0$ and $M_{S}\neq 0$ spin 
 $^{3}E_{u}$ states on the lower AP sheet. It should be noted also, 
that, in general, the overlap integral 
$\left\langle \chi _{+}|\chi_{-}\right\rangle$ depends on the 
$z$-component of  magnetic field 
due to the orbital Zeeman coupling, which contributes to the energy of 
the lower sheet of the adiabatic potential. At 
$\beta H_{z}\gg \lambda_{1}$ an expression for the corresponding 
energy acquires a form:
\begin{equation}
\varepsilon _{-}=
\frac{1}{2}\left( \omega _{\alpha }^{2}Q_{\alpha}^{2}+
\omega _{\beta }^{2}Q_{\beta }^{2}\right) -
\sqrt{\left( V_{\alpha}Q_{\alpha }\right) ^{2}+
\left( V_{\beta }Q_{\beta }\right) ^{2}+
\left(\beta H_{z}g_{1}\right) ^{2}}.
\end{equation}

\section{Conclusions}

We have presented a detailed analysis of the pseudo-Jahn-Teller effect
within a bare electronic $(^{1}A_{1g},^{1,3}E_{u})$ manifold for the 
hole or electron $CuO_4$ centers in doped cuprates. Above we did not 
consider a number of problems generated by the occurrence of the PJT 
centers. Firstly, one should note strong inter-center coupling effects 
caused by the common oxygen ion and the related effects of the 
coupling with either tilting or buckling modes. The PJT centers are 
responsible for the numerous effects of the short-range or long-range 
cooperative PJT ordering observed for the cuprates: some of them have 
been considered earlier \cite{Ivanov,MOK}.  Note Ref.\cite{Ivanov}, 
where some effects of the cooperative PJT ordering are considered 
within a model that could be readily modified for our scenario. 
Authors have performed a model calculation of the specific heat, the 
elastic moduli and the thermal expansion coefficient for the cuprates 
with cooperative PJT effect accompanied by the strong fluctuations of  
crystalline fields.

An important problem is associated with an influence of the PJT
centers to the local boson kinetics and superconductivity. As a whole, 
this is an item for separate discussion though some effects of the 
vibronic reduction and isotope shift were briefly considered earlier 
\cite{isotope}.

In conclusion, we would like to pick up and shortly list again a 
number of experimental data which confirm namely an above developed 
specific scenario of the PJT centers in cuprates:

1. Appearance of the MIR absorption bands for all considered cuprates.

2. The numerous NQR-NMR reveal of the singlet-triplet near-degeneracy 
within ground state manifold for the hole centers.

3. Observation of the anomalously strong anharmonic low temperature
thermal motion of the copper atoms within the copper-oxygen hybrid 
$Q_{e_u}$ mode displayed by the maximum entropy method.

4. Observation of the ferroelectric anomalies.

5. Unusual copper isotope-shift effect in superconducting cuprates.

6. Observation of the specific phonon anomalies.

Among diverse peculiarities of the singlet-triplet PJT centers it 
should be especially emphasized a possible appearance of the so-called 
"tunnel paramagnetic centers" \cite{FTT} which spin states $S=1,M_S 
=\pm 1$ and $S=1,M_S =0$, respectively, are localized within different 
wells of the adiabatic potential. In other words, different spin 
states correspond to different local distortions of the $CuO_4$ 
cluster. Spin dynamics and relaxation for the tunnel paramagnetic 
centers are cruiciably dependent on the magnitude and orientation of  
external magnetic field. These centers could be relatively readily 
transferred to the metastable state. An occurrence of the tunnel 
paramagnetic centers inside the small droplets of the PJT center phase 
was considered \cite{FTT} as an origin of the magnetization and 
magnetostriction anomalies in cuprate $CuO$. Perhaps, these are 
responsible for the unusual magnetic resonance signals observed in 
cuprate $Eu_{2}CuO_4$ \cite{Oseroff}.

It should be noted that we did not undertake the task of reviewing all 
the available experimental data confirming the PJT nature of the 
$CuO_4$ clusters in doped cuprates and comparing them with our model 
approach: it is a separate problem. At the same time, it should be 
noted that the PJT center model is entirely based on the vast amount 
of the experimental material.

We consider the above results as an essential first step to the 
elaboration of the comprehensive theory of the PJT lattice in the 
doped cuprates.

\section{Acknowledgments}

We thank Drs. M.V. Eremin, B.I. Kochelaev, Yu.V. Yablokov, S.Yu. 
Shashkin, A.E. Nikiforov, A.S. Ovchinnikov, V.Ya. Mitrofanov and A.Ya. 
Fishman for stimulating discussions.

\newpage

\begin{table} 
\caption{}
{
Expressions of the decomposition coefficients for electronic 
wave function, extremum coordinates of AP surface and quadratic form 
of AP surface near the extremal points. The following notations are used: 
$q_i^{(0)} = V_i / \omega_i^2$, 
$c_\sigma = \sqrt{\frac{a_\sigma}{a_\sigma + b}}$, 
$d_\sigma = \sqrt{\frac{b}{a_\sigma + b}}$,
$q_1 =(q_x + q_y)/\sqrt{2}$, 
$q_2=(-q_x + q_y)/\sqrt{2}$. 
For the NJT extremum $\varepsilon_0^\sigma = -\Delta - E_{JT}^z$; 
for the JT$_\sigma$ extrema: $\varepsilon_0^\sigma = -E_{JT}^\sigma$; 
for the PJT$_\sigma$ extrema: 
$\varepsilon_0^\sigma = -E_{JT}^\sigma - 
\frac{1}{2} c_\sigma^2 a_\sigma$.
}
\label{table1}
\begin{tabular}{||c||c||c|c|c||c|c|c|c|c||c|}
Type & N & $z$ & $x$ & $y$ & $Q_z^0$ & $Q_\alpha^0$ 
     & $Q_\beta ^0$ & $Q_x^0$ & $Q_y^0$ 
     & $2 \varepsilon(q) - 2 \varepsilon_0^\sigma - 
       \sum_i \omega_i^2 q_i^2$\\ \hline

NJT &1& 1 & 0 & 0 & $-q_z^{(0)}$ & 0 & 0 & 0 & 0 
    & $-\omega_e^2 \eta_z \left( q_x^2+q_y^2 \right)$ \\ \hline

JT$_\alpha$ &1& 0 & 0 & 1 & 0 & $q_\alpha ^{(0)}$ & 0 & 0 & 0 
            & $-\omega_e^2 \kappa_\alpha q_y^2 -
               \omega_\beta^2 \lambda_\alpha q_\beta^2$ \\ \hline
  
            &2& 0 & 1 & 0 & 0 & $-q_\alpha ^{(0)}$ & 0 & 0 & 0 
            & $-\omega_e^2 \kappa_\alpha q_x^2 -
               \omega_\beta^2 \lambda_\alpha q_\beta^2$\\ \hline

JT$_\beta$  &1& 0 & $\frac{-1}{\sqrt{2}}$ & $\frac{1}{\sqrt{2}}$ 
            & 0 & 0 & $q_\beta ^{(0)}$ & 0 & 0 
            & $-\omega_e^2 \kappa_\beta q_2^2 -
               \omega_\alpha^2 \lambda_\beta q_\alpha^2$ \\ \hline

            &2& 0 & $\frac{1}{\sqrt{2}}$ & $\frac{1}{\sqrt{2}}$ 
            & 0 & 0 & $-q_\beta ^{(0)}$ & 0 & 0 
            & $-\omega_e^2\kappa_\beta q_1^2-
               \omega_\alpha ^2\lambda_\beta q_\alpha ^2$\\ \hline

PJT$_\alpha$ &1& $c_\alpha$  & 0 & $d_\alpha$  & $-q_z^{(0)}c_\alpha^2$ 
             & $q_\alpha^{(0)}d_\alpha ^2$ & 0 & 0 
             & $-q_e^{(0)} 2 c_\alpha d_\alpha$ 
             & $\begin{array}{l}
              -\left( \omega_e \upsilon_\alpha q_x +
               \omega_\beta \rho_\alpha q_\beta \right)^2 -\\
              -\left( \omega_z \tau_\alpha q_z +
               \omega_\alpha \mu_\alpha q_\alpha - 
               \omega_e \nu_\alpha q_y \right)^2
               \end{array}$ \\ \hline

             &2& $c_\alpha$ & $d_\alpha$ & 0 & $-q_z^{(0)}c_\alpha^2$ 
             & $-q_\alpha^{(0)} d_\alpha^2$ & 0 
             & $-q_e^{(0)} 2 c_\alpha d_\alpha$ & 0 
             & $\begin{array}{l}
                -\left( \omega_e \upsilon_\alpha q_y +
                \omega_\beta \rho_\alpha q_\beta \right)^2 -\\
                -\left( \omega_z \tau_\alpha q_z -
                \omega_\alpha \mu_\alpha q_\alpha -
                \omega_e \nu_\alpha q_x \right)^2
                \end{array}$ \\ \hline

             &3& $c_\alpha$ & 0 & $-d_\alpha$ & $-q_z^{(0)} c_\alpha^2$
             & $q_\alpha^{(0)} d_\alpha^2$ & 0 & 0 
             & $q_e^{(0)} 2 c_\alpha d_\alpha$ 
             & $\begin{array}{l}
               -\left( \omega_e \upsilon_\alpha q_x -
                \omega_\beta \rho_\alpha q_\beta \right)^2 -\\
               -\left( \omega_z \tau_\alpha q_z +
                \omega_\alpha \mu_\alpha q_\alpha +
                \omega_e \nu_\alpha q_y \right)^2
                \end{array}$ \\ \hline

             &4& $c_\alpha$ & $-d_\alpha$ & 0 & $-q_z^{(0)}c_\alpha^2$
             & $-q_\alpha^{(0)} d_\alpha^2$ & 0 
             & $q_e^{(0)} 2 c_\alpha d_\alpha$ & 0 
             & $\begin{array}{l}
               -\left( \omega_e \upsilon_\alpha q_y -
                \omega_\beta \rho_\alpha q_\beta \right)^2 -\\
               -\left( \omega_z \tau_\alpha q_z -
                \omega_\alpha \mu_\alpha q_\alpha +
                \omega_e \nu_\alpha q_x \right)^2
                \end{array} $ \\ \hline

PJT$_\beta$ &1& $c_\beta$ & $\frac{-d_\beta}{\sqrt{2}}$ 
            & $\frac{d_\beta}{\sqrt{2}}$  
            & $-q_z^{(0)}c_\beta^2$ & 0 & $q_\alpha^{(0)}d_\beta^2$ 
            & $q_e^{(0)}\sqrt{2} c_\beta d_\beta$ 
            & $-q_e^{(0)}\sqrt{2} c_\beta d_\beta$ 
            & $\begin{array}{l}
              -\left( \omega_e \upsilon_\beta q_1 -
               \omega_\alpha \rho_\beta q_\alpha \right)^2 -\\
              -\left( \omega_z \tau_\beta q_z +
               \omega_\beta \mu_\beta q_\beta -
               \omega_e \nu_\beta q_2 \right)^2
               \end{array} $ \\ \hline

            &2& $c_\beta$  & $\frac{d_\beta}{\sqrt{2}}$ 
            & $\frac{d_\beta}{\sqrt{2}}$ & $-q_z^{(0)}c_\beta^2$ 
            & 0 & $-q_\alpha^{(0)} d_\beta^2$ 
            & $-q_e^{(0)}\sqrt{2} c_\beta d_\beta$ 
            & $-q_e^{(0)}\sqrt{2} c_\beta d_\beta$  
            & $\begin{array}{l}
              -\left( \omega_e \upsilon_\beta q_2 - 
               \omega_\alpha \rho_\beta q_\alpha \right)^2 -\\
              -\left( \omega_z \tau_\beta q_z -
               \omega_\beta \mu_\beta q_\beta -
               \omega_e \nu_\beta q_1 \right)^2
               \end{array} $ \\ \hline

            &3& $c_\beta$ & $\frac{d_\beta}{\sqrt{2}}$ 
            & $\frac{-d_\beta}{\sqrt{2}}$ & $-q_z^{(0)} c_\beta^2$ 
            & 0 & $q_\alpha^{(0)} d_\beta^2$ 
            & $-q_e^{(0)}\sqrt{2} c_\beta d_\beta$
            & $q_e^{(0)}\sqrt{2} c_\beta d_\beta$  
            & $\begin{array}{l}
               -\left( \omega_e \upsilon_\beta q_1 +
               \omega_\alpha \rho_\beta q_\alpha \right)^2 -\\
               -\left( \omega_z \tau_\beta q_z +
               \omega_\beta \mu_\beta q_\beta +
               \omega_e \nu_\beta q_2 \right)^2
               \end{array} $ \\ \hline
 
            &4& $c_\beta$ & $\frac{-d_\beta}{\sqrt{2}}$ 
            & $\frac{-d_\beta}{\sqrt{2}}$ & $-q_z^{(0)} c_\beta^2$
            & 0 & $-q_\alpha^{(0)} d_\beta^2$ 
            & $q_e^{(0)}\sqrt{2} c_\beta d_\beta$
            & $q_e^{(0)}\sqrt{2} c_\beta d_\beta$  
            & $\begin{array}{l}
               -\left( \omega_e \upsilon_\beta q_2 +
               \omega_\alpha \rho_\beta q_\alpha \right)^2 -\\
               -\left( \omega_z \tau_\beta q_z - 
               \omega_\beta \mu_\beta q_\beta +
               \omega_e \nu_\beta q_1 \right)^2
               \end{array} $ \\
\end{tabular}
\end{table}

\begin{table}
\caption{}
{
The overlap and tunnel Hamiltonian matrix elements. 
The following notations are used: 
$\sigma_0 = q_\sigma^{(0)} d_\sigma^2$, 
$\sigma_0^{\prime} = q_{\sigma^{\prime }}^{(0)} d_\sigma^2$,
$\gamma_0 = q_e^{(0)} 2 c_\sigma d_\sigma$,
\\
$q_1 = \gamma_0 \cos \psi - \sigma_0 \sin \psi$,
$q_2 = \gamma_0 \sin \psi + \sigma_0 \cos \psi$,
\\
$\Omega_1 = 
\omega_{+} \cos^2 \varphi +
\omega_{-} \sin^2 \varphi$,
$\Omega_1^{\prime} =
\omega_{+}^{\prime} \cos^2 \psi +
\omega_{-}^{\prime} \sin^2 \psi $,
\\
$\Omega_2=
\omega_{+} \sin^2 \varphi +
\omega_{-} \cos^2 \varphi $,
$\Omega_2^{\prime} = 
\omega_{+}^{\prime} \sin^2 \psi +
\omega_{-}^{\prime} \cos^2 \psi$,
\\
$\Omega_{3}=
(-\omega_{+} + \omega_{-}) \sin \varphi \cos \varphi$,
$\Omega_3^{\prime} = 
(-\omega_{+}^{\prime} + \omega_{-}^{\prime}) \sin \psi \cos\psi$,
\\
$\cos \varphi =
\frac{1}{\sqrt{2}} \sqrt{1+\frac{A-B}{\sqrt{(A-B)^2+4C^2}}}$,
$\sin \varphi =
\frac{1}{\sqrt{2}} \sqrt{1-\frac{A-B}{\sqrt{(A-B)^2+4C^2}}}$,
\\
$\cos \psi =
\frac{1}{\sqrt{2}}\sqrt{1+\frac{D-E}{\sqrt{(D-E)^2+4F^2}}}$,
$\sin \psi =
\frac{1}{\sqrt{2}}\sqrt{1-\frac{D-E}{\sqrt{(D-E)^2+4F^2}}}$.
}
\label{table3}

\begin{tabular}{ll}
\multicolumn{2}{l}{a)\quad $H=V_{el}+ T_Q + U_Q +V_{vib}$} \\ \hline
$V_{el}$ & $c_\sigma^2 \ \Delta$ \\ \hline

$T_Q$ & $\frac{1}{4} \left( \omega_{+} + \omega_{-} + 
\omega_{+}^{\prime} + \omega_{-}^{\prime} \right)$ \\ \hline

$U_Q$ & $\frac{1}{4} \Bigl( \frac{\Omega_2}{\omega_{+}\omega_{-}}+
\frac{\Omega_2^{\prime}}{\omega_{+}^{\prime}\omega_{-}^{\prime}} \Bigr)+
\frac{\omega_{\sigma}^2 
\Omega_1^{\prime}}{4\omega_{+}^{\prime}\omega_{-}^{\prime }}+
\frac{\omega_{\sigma^{\prime }}^2 \Omega_1}{4\omega_{+}\omega_{-}} +
\frac{\omega_e^2}{2} \gamma_0^2 +
\frac{\omega_\sigma^2}{2}\sigma_0^2$
\\ \hline

$V_{vib}$ & $-\omega_e^2 \gamma_0^2 -\omega_\sigma^2 \sigma_0^2$ \\ \hline
\multicolumn{2}{l}{b)\quad $S_q =S_q^{el} S_q^n$ 
and $H_q = V_{el}^q + T_Q^q + U_Q^q + V_{vib}^q$} \\ \hline

$S_q^{el}$ & $c_\sigma^2$ \\ \hline

$S_q^n$ & $2\Bigl[ 
\frac{\omega_{+}\omega_{-}
\omega_{+}^{\prime}\omega_{-}^{\prime}}{\left(\omega_{+}\omega_{-} 
+ \Omega_2 \Omega_1^{\prime} \right) 
\left( \omega_{+}^{\prime} \omega_{-}^{\prime} 
+ \Omega_1 \Omega_2^{\prime} \right) }\Bigr]^{1/2}
\exp \left\{ -\frac{\omega_{+} \omega_{-} 
\left( \omega_{+}^{\prime } q_1^2 +
\omega_{-}^{\prime} q_2^2 \right) +
\omega_{+}^{\prime} \omega_{-}^{\prime} 
\Omega_2 \sigma_0^2}{ \omega_{+} \omega_{-} +
\Omega_2 \Omega_1^{\prime}} \right\}$ \\ \hline

$V_{el}^q$ & $c_\sigma^2 \ \Delta \ S_q^n$ \\ \hline

$T_Q^q$ & $
\begin{array}{l}
S_q \Bigl[ \frac{1}{2} 
\left( \omega_{+} + \omega_{-} + 
\omega_{+}^{\prime } + \omega_{-}^{\prime} \right) -
\frac{\Omega_2^2 (A+D) + \Omega_3^2 B + 
2 \Omega_2 \Omega_3 C}{2\left( \omega_{+} \omega_{-} + \Omega_2 \Omega_1^{\prime} \right)^2}
\left( \gamma_0 \Omega_2^{\prime} + 
\sigma_0 \Omega_3^{\prime}\right)^2 -  \\ 

-\frac{1}{4} \left( \frac{(\Omega_1 + \Omega_1^{\prime}) B +
\Omega_2(A+B) + 2 \Omega_3 C}{\omega_{+} \omega_{-} + 
\Omega_2 \Omega_1^{\prime}} + \frac{(\Omega_1 + 
\Omega_1^{\prime}) E + \Omega_2^{\prime}(A+B) + 
2\Omega_3^{\prime} F}{\omega_{+}^{\prime} \omega_{-}^{\prime} +
\Omega_1 \Omega_2^{\prime}} \right) - 
\frac{\Omega_2 (\gamma_0 \Omega_2^{\prime} +
\sigma_0\Omega_3^{\prime}) (\sigma_0 F - \gamma_0 D)}{\omega_{+} \omega_{-} + 
\Omega_2 \Omega_1^{\prime}} - \frac{D}{2} \gamma_0^2-
\frac{E}{2} \sigma_0^2 + F \sigma_0 \gamma_0 \Bigr]
\end{array}
$ \\ \hline

$U_Q^q$ & $S_q \Bigl[ \frac{1}{4} 
\bigl( \frac{2\omega_e^2 \Omega_2 +
\omega_\beta^2 (\Omega_1 + 
\Omega_1^{\prime})}{\omega_{+} \omega_{-} + 
\Omega_2 \Omega_1^{\prime}} + 
\frac{2 \omega_e^2 \Omega_2^{\prime} + 
\omega_\sigma^2 (\Omega_1 + 
\Omega_1^{\prime})}{\omega_{+}^{\prime} \omega_{-}^{\prime} + 
\Omega_1 \Omega_2^{\prime}} \bigr) + 
\frac{2 \omega_e^2 \Omega_2^2 + 
\omega_{\sigma^{\prime }}^2 
\Omega_3^2}{2 \left( \omega_{+} \omega_{-} + 
\Omega_2 \Omega_1^{\prime} \right)^2}
(\gamma_0 \Omega_2^{\prime} +
\sigma_0 \Omega_3^{\prime})^2 \Bigr] $ \\ 
\hline

$V_{vib}^q$ & $S_q^n \left[-\omega_e^2 \Omega_2 \gamma_0 + 
\omega_{\sigma^{\prime}}^2 \Omega_3 \sigma_0^{\prime} \right] 
\frac{\gamma_0 \Omega_2^{\prime} + 
\sigma_0 \Omega_3^{\prime}}{\omega_{+} \omega_{-} + 
\Omega_2 \Omega_1^{\prime}}$ \\ \hline
\multicolumn{2}{l}{c)\quad $S_d = S_d^{el} S_d^n$ 
and $H_d = V_{el}^d + T_Q^d + U_Q^d + V_{vib}^d$ } \\ \hline

$S_d^{el}$ & $c_\sigma^2 - d_\sigma^2$ \\ \hline 

$S_d^n$ & $\bigl[ \frac{\omega_{+} \omega_{-} \omega_{+}^{\prime} 
\omega_{-}^{\prime }}{\Omega_1 \Omega_2 
\Omega_1^{\prime} \Omega_2^{\prime}} \bigr]^{1/2}
\exp \bigl\{ -\frac{\omega_{+}^{\prime} 
\omega_{-}^{\prime}}{\Omega_2^{\prime}} 
\gamma_0^2 \bigr\} $ \\ \hline

$V_{el}^d$ & $c_\sigma^2 \ \Delta \ S_d^n$ \\ \hline

$T_Q^d$ & $S_d^n \bigl[ \frac{1}{2}
( \omega_{+} + \omega_{-} + 
\omega_{+}^{\prime}+\omega_{-}^{\prime} ) -
\frac{1}{4} \bigl( \frac{A}{\Omega_1} + \frac{B}{\Omega_2} +
\frac{D}{\Omega_1^{\prime}} + \frac{E}{\Omega_2^{\prime}} \bigr)- \frac{\omega_{+}^{\prime \ 2} 
\omega_{-}^{\prime \ 2}}{2 \Omega_2^{\prime \ 2}}
\gamma_0^2 \bigr] $ \\ \hline

$U_Q^d$ & $S_d \bigl[ \frac{1}{4} 
\bigl( \frac{\omega_e^2}{\Omega_1} + 
\frac{\omega_e^2}{\Omega_1^{\prime }} + 
\frac{\omega_\sigma^2}{\Omega_2^{\prime}} + 
\frac{\omega_{\sigma^{\prime}}^2}{\Omega_2} \bigr) +
\frac{\omega_{\sigma }^2}{2} \sigma_0^2 \bigr]$ \\ \hline

$V_{vib}^d$ & $S_d^n \omega_\sigma^2 \sigma_0 
\bigl( \sigma_0 + 
\frac{\Omega_3^{\prime}}{\Omega_2^{\prime }}\gamma_0 \bigr)$ \\
\end{tabular}
\end{table}

\begin{table} 
\caption{ 
The reduction factors: the 
$\tilde{\Gamma}$; $\Gamma $, $\Gamma ^{\prime }$;
$\Gamma _{1}$, $\Gamma _{1}^{\prime }$ label 
the transformation properties of the
electronic operator (with 
$\Sigma =\alpha $ at $\Sigma ^{\prime }=\beta $, and
$\Sigma =\beta $ at $\Sigma ^{\prime }=\alpha $), 
the vibronic wave function and the
electronic wave function, respectively. Additional notations:
$\left\langle \chi _{a_{1g}}^{2}\right\rangle =
1+2S_{q}^{n}+S_{d}^{n}$,
$\left\langle \chi _{\sigma }^{2}\right\rangle =
1-2S_{q}^{n}+S_{d}^{n}$,
$\left\langle \chi _{e_{u}}^{2}\right\rangle =
1-S_{d}^{n}$.
}
\label{table4}
\begin{tabular}{|l|l|l|l|l|c|}
$\tilde{\Gamma}$ & $\Gamma$ & $\Gamma^{\prime}$ 
& $\Gamma_1$ & $\Gamma_1^{\prime}$ 
& $K_{\tilde{\Gamma}} 
\left( 
\begin{array}{ll}
\Gamma  & \Gamma^{\prime} \\ 
\Gamma_1 & \Gamma_1^{\prime}
\end{array}
\right) $ \\ \hline

$A_{1g}$ & $A_{1g}$ & $A_{1g}$ & $A_{1g}$ & $A_{1g}$ 
& $ N_{A_{1g}}^{-2}\ c_\sigma^2 \ 
\left\langle \chi_{a_{1g}}^2 \right\rangle $ \\ \hline

& $A_{1g}$ & $A_{1g}$ & $E_u$ & $E_u$ 
& $2^{-1/2}\ N_{A_{1g}}^{-2}\ d_\sigma^2 \ 
\left\langle \chi_{e_u}^2 \right\rangle $ \\ \hline

& $E_u$ & $E_u$ & $A_{1g}$ & $A_{1g}$ 
& $2^{1/2} \ N_{E_u}^{-2} \ c_\sigma^2 \ 
\left\langle \chi_{e_u}^2 \right\rangle $ \\ \hline

& $E_u$ & $E_u$ & $E_u$ & $E_u$ 
& $2^{-1}\ N_{E_u}^{-2} \ d_\sigma^2 \ 
\left\{ \left\langle \chi_{a_{1g}}^2 \right\rangle +
\left\langle \chi_{e_u}^2 \right\rangle\right\} $ \\ \hline

& $\Sigma$ & $\Sigma$ & $A_{1g}$ & $A_{1g}$ 
& $N_\Sigma^{-2} \ c_\sigma^2 \ 
\left\langle \chi_\sigma^2 \right\rangle $ \\ \hline

& $\Sigma$ & $\Sigma$ & $E_u$ & $E_u$ 
& $2^{-1/2} \ N_\Sigma^{-2} \ d_\sigma^2 \ 
\left\langle \chi_{e_u}^2 \right\rangle $ \\ \hline

$\Sigma$ & $E_u$ & $E_u$ & $E_u$ & $E_u$ 
& $2^{-1} \ N_{E_u}^{-2} \ d_\sigma^2 \ 
\left\{ \left\langle \chi_{a_{1g}}^2 \right\rangle +
\left\langle \chi_{e_u}^2 \right\rangle\right\} $ \\ \hline

& $A_{1g}$ & $\Sigma$ & $E_u$ & $E_u$ 
& $-2^{-1/2} \ N_{A_{1g}}^{-1} \ N_\Sigma^{-1} \ d_\sigma^2 \ 
\left\langle \chi_{e_u}^2 \right\rangle$ \\ \hline

$\Sigma^{\prime}$ & $E_u$ & $E_u$ & $E_u$ & $E_u$ 
& $2^{-1} \ N_{E_u}^{-2} \ d_\sigma^2 \ \left\{
\left\langle \chi_{a_{1g}}^2 \right\rangle -
\left\langle \chi_{e_u}^2 \right\rangle \right\} $ \\ \hline

$A_{2g}$ & $E_u$ & $E_u$ & $E_u$ & $E_u$ 
& $2^{-1} \ N_{E_u}^{-2} \ d_\sigma^2 \ \left\{ 
\left\langle \chi_{a_{1g}}^2 \right\rangle -
\left\langle \chi_{e_u}^2 \right\rangle\right\} $ \\ \hline

$E_u$ & $A_{1g}$ & $E_u$ & $A_{1g}$ & $E_u$ 
& $2^{-1/2} \ N_{A_{1g}}^{-1} \ N_{B_{1g}}^{-1} \ c_\sigma \ d_\sigma \
\left\langle \chi_{a_{1g}}^2 \right\rangle $ \\ \hline

& $A_{1g}$ & $E_u$ & $E_u$ & $A_{1g}$ 
& $2^{-1/2} \ N_{A_{1g}}^{-1} \ N_{B_{1g}}^{-1} \ c_\sigma \ d_\sigma \
\left\langle \chi_{e_u}^2 \right\rangle $ \\ \hline

& $\Sigma$ & $E_u$ & $A_{1g}$ & $E_u$ 
& $-2^{-1/2} \ N_\Sigma^{-1} \ N_{E_u}^{-1} \ c_\sigma \ d_\sigma \
\left\langle \chi_\sigma^2 \right\rangle $ \\ \hline

& $\Sigma$ & $E_u$ & $E_u$ & $E_u$ 
& $-2^{-1/2} \ N_\Sigma^{-1} \ N_{E_u}^{-1} \ c_\sigma \ d_\sigma \
\left\langle \chi_{e_u}^2 \right\rangle $ \\
\end{tabular}
\end{table}

\newpage
{\Large Figure captions}

\begin{figure} 
\caption{Correlation effects in the energy spectrum of the basic 
$CuO_{4}^{6-}$ cluster and the hole $CuO_{4}^{5-}$ center with 
numerical values (in eV) typical for oxides like $CuO$. On the right 
hand side we show a formation of the fundamental absorption spectra 
for the parent and hole doped oxides with peculiar MIR band in the 
latter case.The lower panel is an illustration of the bare (a) 
$CuO_{4}^{5-}$ center energy spectrum modification with taking into 
account the strong pseudo-Jahn-Teller-effect (b) and the tunnel 
splitting of the ground vibronic states (c). Distortions of the 
$CuO_4$ cluster assotiated with the different AP minima are shown 
schematically in the insert.}
\label{fig1}
\end{figure}

\begin{figure} 
\caption{ Possible distortions of the $CuO_{4}$ cluster in the a)
JT$_{\alpha }$ minimum; b) JT$_{\beta }$ minimum; c) PJT$_{\alpha }$ 
minimum (the dipole moment is oriented along the $CuO_{4}$ cluster 
 diagonal); b) PJT$_{\beta}$ minimum (the rotation angle value for the 
dipole moment with respect to that in PJT$_{\alpha }$ minima is 
$\phi = \pi /4$). }
\label{fig2}
\end{figure}

\begin{figure} 
\caption{ The allocation of the PJT$_\sigma$ minima in a space of 
symmetric coordinates $Q_\sigma,Q_1,Q_2$ ($\sigma$ is the 
''strong'' rhombic mode). If $\sigma =\alpha $, then $Q_1=Q_x$, 
$Q_2=Q_y$; if $\sigma=\beta$, then 
$Q_1=\frac{Q_x+Q_y}{2}$, $Q_2=\frac{-Q_x+Q_y}{2}$.  }
\label{fig3}
\end{figure}

\begin{figure} 
\caption{ 
The diagram of the lower AP sheet states in the
$(\Delta,E_{JT}^\sigma,E_{JT}^e)$ space at certain fixed
$E_{JT}^z=E_{JT}^{z,0}$ value. The planes $a_\sigma=0$ and $b=0$ 
divide the $(\Delta ,E_{JT}^\sigma,E_{JT}^e)$ space into four 
parts corresponding to NJT-, JT$_\sigma$-, PJT$_\sigma$- or 
NJT+JT$_\sigma$-type of the lower AP sheet minima. In the 
NJT+JT$_\sigma$ region the plane is shown, where the NJT- and 
JT$_\sigma$ minimum energies are equal. }
\label{fig4}
\end{figure}

\begin{figure} 
\caption{
a) The diagram of states of the lower AP sheet on the 
$(\Delta,E_{JT}^\sigma)$ plane at certain fixed $E_{JT}^z$ and 
$E_{JT}^e$ value. The boundaries $a_\sigma=0$ and $b=0$ of the 
regions with different type of the lower AP sheet minima are shown. 
The lines $E_{JT}^\sigma = E_{JT}^{\sigma ,1}$ and 
$E_{JT}^\sigma =E_{JT}^{\sigma,2}$ correspond to the two different 
relative intensities of the vibronic coupling via $\sigma$ mode 
and $e_u$ mode. b) The diagram of states of the upper AP sheets on 
the $(\Delta,E_{JT}^\sigma)$ plane at certain fixed $E_{JT}^z$ and 
$E_{JT}^e$ value. In the region to the left of the 
$\Delta = -2E_{JT}^z$ line there is NJT minimum on the upper AP sheet 
and only the trivial non-analytical minimum on the middle AP sheet. 
In the region to the right of the $\Delta =2E_{JT}^\sigma$ line there 
is $JT_\sigma$ minima on the middle AP sheet and the trivial 
non-analytical minimum on the upper AP sheet. At 
$E_{JT}^z<2E_{JT}^e$ the regions of parameters exist, where the 
PJT$_\sigma$ minima on the lower AP sheet coexist with the NJT 
minimum on the upper sheet or with the JT$_\sigma$ minima on the 
middle sheet. Between the $\Delta =-2E_{JT}^z$ and 
$\Delta=2E_{JT}^\sigma$ lines only the trivial non-analytical 
minima on the upper sheets exist. }
\label{fig5}
\end{figure}

\begin{figure} 
\caption{ 
The $\Delta$ dependencies of a) the lower tunnel energy levels 
and b) the corresponding tunnel splittings at $k_e=3$, 
$k_\sigma/k_e=0.5$, $k_{\sigma^{\prime}}/k_\sigma = 0.5$, $k_z=0$ 
where $k_{i}^2=2E_{JT}^i/\hbar \omega_i$. A significant 
increasing of the $\Sigma$ level energy at the $\Delta$ interval 
border is connected with non-applicability of the chosen basis for a 
description of the exited vibronic states in the situation of shallow 
minima.}
\label{fig6}
\end{figure}

\begin{figure} 
\caption{
The possible $CuO_4$ cluster distortions in the equipotential
continuum of minima as function of $\varphi $ angle (see text). The 
$n$ points correspond to the ion allocations at $\varphi =n\pi /2$. 
The a) and b) panels differ by the initial mutual orientation of the 
$b_{1g}$ and $b_{2g}$ distortions. The top and bottom fragments on 
each panel correspond to different initial mutual orientation of the 
($b_{1g},b_{2g}$) and $e_u$ distortions. The possible trajectories 
of the oxygen ions for the different values of 
$\left|2\frac{q_e^{(0)}}{q_\sigma^{(0)}}
\left( \frac{a_\sigma}{b} \right)^{(1/2)} \right|$ are shown. 
The radius of the copper ion trajectory is proportional to 
$q_e^{(0)}$. The equivalent $CuO_4$ cluster 
distortions with opposite phase of the copper ion 
($n_{Cu}\rightarrow n_{Cu}+4$) are also possible.}
\label{fig7}
\end{figure}

\begin{figure} 
\caption{
The reduction factor dependencies 
a) as a function of $\Delta$ 
at $k_e=3$, $k_\sigma / k_e = 0.5$, 
$k_{\sigma^{\prime}}/k_\sigma = 0.5$, $k_z=0$ 
($k_i^2=2E_{JT}^i/\hbar \omega_i$); 
b) as a function of dimensionless coupling constant $k$ at 
$k_e=k$, $k_\sigma/k_e=0.5$, 
$k_{\sigma^{\prime}}/k_\sigma = 0.5$, $k_z=0$.
}
\label{fig8}
\end{figure}

\end{document}